\documentclass[showpacs,prc,preprint,nofootinbib,showkeys,unsortedaddress]{revtex4}
\pdfoutput=1

\usepackage{bm}
\usepackage{amsmath}
\usepackage{graphicx}
\usepackage{subfigure}
\usepackage[colorlinks=true,linktocpage=true,linkcolor=blue,citecolor=blue]{hyperref}
\usepackage[usenames,dvipsnames]{color}

\def\be{\begin{equation}}
\def\ee{\end{equation}}

\def\ba{\begin{eqnarray}}
\def\ea{\end{eqnarray}}

\definecolor{darkblue}{RGB}{0,0,160}

\newcommand{\checked}[1]{}

\begin{document}

\title{Bulk viscous evolution within anisotropic hydrodynamics}

\author{Mohammad Nopoush} 
\affiliation{Department of Physics, Kent State University, Kent, OH 44242 United States}

\author{Radoslaw Ryblewski}
\affiliation{Department of Physics, Kent State University, Kent, OH 44242 United States}
\affiliation{The H. Niewodnicza\'nski Institute of Nuclear Physics, Polish Academy of Sciences, PL-31342 Krak\'ow, Poland} 

\author{Michael Strickland} 
\affiliation{Department of Physics, Kent State University, Kent, OH 44242 United States}

\begin{abstract}
We derive a system of moment-based dynamical equations that describe the 1+1d space-time evolution of a cylindrically symmetric massive gas undergoing boost-invariant longitudinal expansion.  Extending previous work, we introduce an explicit degree of freedom associated with the bulk pressure of the system.  The resulting form generalizes the ellipsoidal one-particle distribution function appropriate for massless particles to massive particles.  Using this generalized form, we obtain a system of partial differential equations that can be solved numerically.  In order to assess the performance of this scheme, we compare the resulting anisotropic hydrodynamics solutions with the exact solution of the 0+1d Boltzmann equation in the relaxation time approximation.  We find that the inclusion of the bulk degree of freedom improves agreement between anisotropic hydrodynamics and the exact solution for a massive gas.
\end{abstract}

\date{\today}



\maketitle

\section{Introduction}
\label{sect:intro}

The goal of ultrarelativistic heavy-ion collision experiments at the Relativistic Heavy Ion Collider at Brookhaven National Laboratory (RHIC) and the Large Hadron Collider (LHC) at CERN is to create a tiny volume of matter ($\sim$ 100\,$-$\,1000 fm$^3$) that has been heated to a temperature exceeding that necessary to create a quark-gluon plasma (QGP), namely $T \gtrsim 175 \; \rm MeV$.  Early on it was shown that ideal relativistic hydrodynamics was able to reproduce the soft collective flow of the QGP and hadronic spectra produced at RHIC reasonably well \cite{Huovinen:2001cy,Hirano:2002ds,Kolb:2003dz}.  Based on this, there was a concerted effort to develop a systematic framework for describing the soft collective motion.  This effort resulted in a number of works dedicated to the development and application of relativistic viscous hydrodynamics to relativistic heavy-ion collisions~\cite{Muronga:2001zk,Muronga:2003ta,Muronga:2004sf,Heinz:2005bw,Baier:2006um,Romatschke:2007mq,Baier:2007ix,Dusling:2007gi,Luzum:2008cw,Song:2008hj,Heinz:2009xj,El:2009vj,PeraltaRamos:2009kg,PeraltaRamos:2010je,Denicol:2010tr,Denicol:2010xn,Schenke:2010rr,Schenke:2011tv,Bozek:2011wa,Niemi:2011ix,Niemi:2012ry,Bozek:2012qs,Denicol:2012cn,Denicol:2012es,PeraltaRamos:2012xk,Calzetta:2014hra,Denicol:2014vaa}.  Part of the motivation for these developments was a calculation within the framework of strongly-coupled ${\cal N}=4$ supersymmetric Yang Mills theory which demonstrated that there exists a lower bound on the ratio of the shear viscosity to entropy density \cite{Kovtun:2004de}.  This development implied that it was necessary to use viscous rather than ideal relativistic hydrodynamics, sparking a tremendous amount of theoretical activity on this front.  This activity seems to have been worth it since comparisons of the resulting viscous hydrodynamical models with experimental results show better agreement between theory and experiment than obtained using simple ideal hydrodynamical models.

However, introducing finite shear viscosity into the relativistic hydrodynamics formalism is not without complications.  One complication is that, for finite shear viscosity to entropy density ratio $\bar\eta \equiv \eta/{\cal S}$, relativistic viscous hydrodynamical simulations predict rather sizable pressure anisotropies in the local rest frame \cite{Martinez:2009mf,Strickland:2013uga,Strickland:2014eua}.  In practice, one finds that the transverse pressure, ${\cal P}_T$, exceeds the longitudinal pressure, ${\cal P}_L$, with the difference being largest at early times $\tau \lesssim$ 2 fm/c.  In addition, the size of the pressure anisotropies increases as one approaches the transverse and longitudinal edges of the system where the particles are approximately free streaming.  Faced with such potentially large momentum-space anisotropies a formalism called anisotropic hydrodynamics (aHydro) was created \cite{Martinez:2010sc,Florkowski:2010cf,Ryblewski:2010bs,Martinez:2010sd,Ryblewski:2011aq,Florkowski:2011jg,Martinez:2012tu,Ryblewski:2012rr,Florkowski:2012as,Bazow:2013ifa,Florkowski:2013uqa,Tinti:2013vba,Florkowski:2014txa,Florkowski:2014bba}.  The basic idea behind this method is to take into account the largest corrections to the ideal fluid form non-perturbatively by subsuming them into the argument of the leading-order one-particle distribution function
\begin{equation}
   f(x,p) = 
   f_{\rm iso}\!\left(\frac{\sqrt{p^\mu \Xi_{\mu\nu}(x)p^\nu}}{\lambda(x)}, \frac{\mu(x)}{\lambda(x)}\right)
   + \delta\!\tilde{f}(x,p) \, .
\label{eq:ahexp}
\end{equation}
Above $\Xi_{\mu\nu}$ is a symmetric second-rank tensor that measures the amount of momentum-space anisotropy, $\lambda$ is a temperature-like scale which can be identified with the true temperature of the system only in the isotropic equilibrium limit, and $\mu(x)$ is the effective chemical potential. The second term $\delta\!\tilde{f}$ collects corrections to the leading-order form and can be treated perturbatively using a systematic expansion in (modified) Knudsen and Reynolds numbers \cite{Bazow:2013ifa}.  The symmetric tensor $\Xi_{\mu\nu}$ can possess both diagonal and off-diagonal terms, however, in practice one finds that the off-diagonal terms are small \cite{Song:2009gc}.  Therefore, a good leading-order approximation might be to include only the large diagonal anisotropies in $\Xi_{\mu\nu}$.  This was the original spirit of anisotropic hydrodynamics; however, in the original aHydro papers it was assumed that $\Xi^{\mu\nu} = {\rm diag}(1,0,0,\xi)$, where $\xi$ was the spheroidal anisotropy parameter \cite{Romatschke:2003ms}, in order to create a tractable scheme to describe the evolution of the 0+1d systems \cite{Martinez:2010sc,Florkowski:2010cf}.  For a system of massless particles, the spheroidal approximation was shown to analytically reproduce both the small and large $\bar{\eta}$ limits \cite{Martinez:2010sc}.  In addition, the spheroidal form was shown to also reproduce second-order Israel-Stewart viscous hydrodynamics in the small-$\xi$ limit \cite{Martinez:2010sc}.

In order to make the formalism more general we would like to extend it to include finite particle masses and ellipsoidal momentum-space anisotropies.  For a system of massless particles that is cylindrically symmetric and boost-invariant, Tinti and Florkowski \cite{Tinti:2013vba} recently demonstrated how to obtain dynamical equations for the diagonal components of $\Xi^{\mu\nu}$ using the first and second moments of the Boltzmann equation.  The resulting dynamical equations reproduce the exact solution to the 0+1d Boltzmann equation \cite{Florkowski:2014bba} for a massless gas better than the original aHydro formulation, which used the zeroth and first moments of the Boltzmann equation \cite{Florkowski:2013lza,Florkowski:2013lya}.  In practice, however, Tinti and Florkowski simply disregarded the equations resulting from the zeroth and $ux$-projected second moments of the Boltzmann equation.  This naturally leads one to ask (i) what is the correct procedure for selecting the necessary dynamical equations if the system is overdetermined (apart from the requirement that the equations reduce to second-order viscous hydrodynamics) and (ii) is it possible to add an additional degree of freedom(s) so that the system is no longer overdetermined.  In addition, in a subsequent work Florkowski, Tinti, and two of the current authors made a naive extension of the ellipsoidal formalism to a massive gas \cite{Florkowski:2014bba}.  The introduction of finite mass breaks the conformality of the system allowing for finite bulk pressures in the system.  In Ref.~\cite{Florkowski:2014bba} a careful comparison of aHydro to exact solutions of the massive 0+1d Boltzmann equation in relaxation time approximation \cite{Florkowski:2014sfa} showed that neither spheroidal nor ellipsoidal aHydro were able to reproduce the early-time dynamics of the bulk pressure.

In this paper, we continue our consideration of aHydro applied to a massive gas in the relaxation time approximation.  We generalize the ellipsoidal formalism employed in Refs.~\cite{Tinti:2013vba,Florkowski:2014bba} to include an explicit degree of freedom that can be associated with the bulk pressure.  This is done in a similar manner to standard viscous hydrodynamical treatments where the viscous correction is decomposed into a traceless and traceful parts associated with shear and bulk corrections, respectively; however, in our case this decomposition is performed at the level of the $\Xi_{\mu\nu}$ tensor.  We derive the dynamical equations necessary to describe a boost-invariant and cylindrically symmetric system of massive particles and then specialize to the case of a transversally homogeneous system so that we can compare with the exact solution of the 0+1d Boltzmann equation in the relaxation time approximation.  In the process we show that for $m>0$ the $uu$-projection of the second moment of the Boltzmann equation is identical to the zeroth moment equation.  As a result, for a 0+1d system the zeroth moment equation can be used to evolve the bulk degree of freedom and the system.  In the general case, the dynamical equations derived herein can be used to describe both the transverse and longitudinal dynamics of a massive gas in the relaxation time approximation.

The structure of the paper is as follows.  In Sec.~\ref{sect:setup} we specify the setup and assumptions used for the four-vector basis for the system and the distribution function ansatz. In Sec.~\ref{sect:bulkvars} we present expressions for the number density, energy density, transverse pressure, and longitudinal pressure using an ellipsoidally deformed distribution which includes the bulk degree of freedom.  In Sec.~\ref{sect:beqmoments} we compute the zeroth, first, and second moments of the Boltzmann equation for a boost-invariant cylindrically symmetric massive gas.  In Sec.~\ref{sect:results} we present comparisons of numerical solution of the aHydro equations for a 0+1d massive gas with the exact solution.  In Sec.~\ref{sect:conc} we present our conclusions and an outlook for the future.  In App.~\ref{app:hfuncs} we collect explicit expressions for and asymptotic expansions of the ${\cal H}$ functions which appear in expressions for the bulk properties.  In App.~\ref{app:derivs} we collect explicit expressions for various derivatives and derivative-projections which appear in the equations of motion.

\section{Setup}
\label{sect:setup}

We begin by specifying the basic setup necessary for treating a boost-invariant and cylindrically symmetric system allowing for an explicit bulk degree of freedom.

\subsection{Vector Basis}
\label{sect:basis}

A general tensor basis can be constructed by introducing four 4-vectors which in the local rest frame (LRF) are
\ba
&&X^\mu_{0,{\rm LRF}} \equiv u^\mu_{\rm LRF} = (1,0,0,0) \, , \nonumber \\
&&X^\mu_{1,{\rm LRF}} \equiv x^\mu_{\rm LRF} = (0,1,0,0) \, , \nonumber \\
&&X^\mu_{2,{\rm LRF}} \equiv y^\mu_{\rm LRF} = (0,0,1,0) \, , \nonumber \\
&&X^\mu_{3,{\rm LRF}} \equiv z^\mu_{\rm LRF} = (0,0,0,1) \, .
\label{eq:rfbasis}
\ea
\checked{rs}
These 4-vectors are orthonormal in all frames.  The vector $X^\mu_0$ is associated with the four-velocity of the local rest frame and is conventionally called $u^\mu$.  One can also identify $X^\mu_1 = x^\mu$, $X^\mu_2 = y^\mu$, and $X^\mu_3 = z^\mu$ as indicated above.  We will use the two different labels for these vectors interchangeably depending on convenience since the notation with numerical indices allows for more compact expressions in many cases.  Note that in the lab frame, the three  spacelike vectors $X^\mu_i$ can be written entirely in terms of $X^\mu_0=u^\mu$.  This is because $X^\mu_i$ can be obtained by a sequence of Lorentz transformations/rotations applied to the local rest frame expressions specified above.

We point out that one can express the metric tensor itself in terms of these 4-vectors as
\be
g^{\mu \nu}= X^\mu_0 X^\nu_0 - \sum_i X^\mu_i X^\nu_i \, ,
\label{eq:gbasis}
\ee
\checked{rs}
where the sum extends over $i=1,2,3$.  In addition, the standard transverse projection operator which is orthogonal to $X^\mu_0$ can be expressed in terms of the basis (\ref{eq:rfbasis})
\be
\Delta^{\mu \nu} = g^{\mu\nu} - X^\mu_0 X^\nu_0 = - \sum_i X^\mu_i X^\nu_i \, ,
\label{eq:transproj}
\ee
\checked{rs}
from which finds $u_\mu \Delta^{\mu \nu} = u_\nu \Delta^{\mu \nu} = 0$ as expected.  We note that the spacelike components of the tensor basis are eigenfunctions of this operator, i.e. $X_{i\mu} \Delta^{\mu \nu} = X^\nu_{i}$.

Following Ref.~\cite{Tinti:2013vba} we begin by considering a cylindrically symmetric system undergoing boost-invariant expansion.  In this case, one can parameterize the basis vectors as

\be
\begin{array}{ll}
\begin{aligned}
u^0 &= \cosh\theta_\perp \cosh\eta_\parallel \, , \\
u^1 &= \sinh\theta_\perp \cos\phi \, , \\
u^2 &= \sinh\theta_\perp \sin\phi \, , \\
u^3 &= \cosh\theta_\perp \sinh\eta_\parallel \, ,
\end{aligned}
\quad \quad \quad &
\begin{aligned}
x^0 &= \sinh\theta_\perp \cosh\eta_\parallel \, , \\
x^1 &= \cosh\theta_\perp \cos\phi \, , \\
x^2 &= \cosh\theta_\perp \sin\phi \, , \\
x^3 &= \sinh\theta_\perp \sinh\eta_\parallel \, ,
\end{aligned}
\\
&\vspace{-3mm}\\
\begin{aligned}
y^0 &= 0 \, , \\
y^1 &= -\sin\phi \, , \\
y^2 &= \cos\phi \, , \\
y^3 &= 0 \, ,
\end{aligned}
\quad \quad  \quad &
\begin{aligned}
z^0 &= \sinh\eta_\parallel \, , \\
z^1 &= 0 \, , \\
z^2 &= 0 \, , \\
z^3 &= \cosh\eta_\parallel \, .
\\
\\
\end{aligned}
\end{array}
\label{eq:expvectorbasis}
\ee
\checked{rs}

For future use, it is convenient to introduce the expansion tensor
\be
\theta^{\mu\nu} \equiv {\Delta^\mu}_{\!\alpha} {\Delta^\nu}_{\!\beta} \, \partial^{(\alpha}u^{\beta)} \, ,
\ee
\checked{rs}
where the parentheses indicate the symmetric part, i.e. $A^{(\alpha\beta)} =  ( A^{\alpha\beta} + A^{\beta\alpha})/2$.  For a boost-invariant cylindrically symmetric system one can decompose the expansion tensor as \cite{Florkowski:2011jg,Tinti:2013vba}
\be
\theta^{\mu\nu} = \theta_x x^\mu x^\nu + \theta_y y^\mu y^\nu + \theta_z z^\mu z^\nu \, ,
\ee
\checked{rs}
where, using the explicit forms of the basis vectors (\ref{eq:expvectorbasis}), one finds
\ba
\theta_x &=& -\frac{\partial\theta_\perp}{\partial r} \cosh\theta_\perp - \frac{\partial\theta_\perp}{\partial \tau} \sinh\theta_\perp \, , \\
\theta_y &=& -\frac{\sinh\theta_\perp}{r} \, , \\
\theta_z &=& -\frac{\cosh\theta_\perp}{r} \, .
\label{eq:thetais}
\ea
\checked{rs}
Finally, we note that the expansion scalar $\theta = \Delta^{\mu\nu} \theta_{\mu\nu} = \partial_\mu u^\mu$ obeys
\ba
\theta = -\theta_x - \theta_y - \theta_z \, .
\ea
\checked{rs}
We list some additional properties such as the derivatives of the basis vectors in App.~\ref{app:derivs}.

\subsection{Ellipsoidal form including bulk pressure degree of freedom}

We start by introducing the anisotropy tensor
\be
\Xi^{\mu\nu} = u^\mu u^\nu + \xi^{\mu\nu} - \Delta^{\mu\nu} \Phi \, ,
\ee
\checked{rs}
where $u^\mu$ is the four-velocity associated with the local rest frame, $\xi^{\mu\nu}$ is a symmetric and traceless tensor, and $\Phi$ is the bulk degree of freedom.  The quantities $u^\mu$, $\xi^{\mu\nu}$, and $\Phi$ are understood to be functions of space and time and obey
\ba
u^\mu u_\mu &=& 1 \, , \\
{\xi^{\mu}}_\mu &=& 0 \, , \\
{\Delta^\mu}_\mu &=& 3 \, , \\
u_\mu \xi^{\mu\nu} &=& u_\mu \Delta^{\mu\nu} = 0 \, ,
\ea
\checked{rs}
and as a result
\be
{\Xi^\mu}_\mu = 1 - 3 \Phi \, .
\ee
\checked{rs}
The anisotropic one-particle distribution function can be constructed using $\Xi^{\mu\nu}$ as
\be
f(x,p) = f_{\rm iso}\!\left(\frac{1}{\lambda}\sqrt{p_\mu \Xi^{\mu\nu} p_\nu}\right) ,
\label{eq:genf}
\ee
\checked{rs}
where $\lambda$ has dimensions of energy and can be identified with the temperature only in the isotropic equilibrium limit ($\xi^{\mu\nu} = 0$ and $\Phi=0$).\footnote{Herein we assume that the chemical potential is zero.}  We note that in practice $f_{\rm iso}$ need not be a thermal equilibrium distribution.  However, unless one expects there to be a non-thermal fixed point at late times, it is appropriate to take $f_{\rm iso}$ to be a thermal equilibrium distribution function of the form
\begin{equation}
f_{\rm iso}(x) = f_{\rm eq}(x) = \Big( e^x + a \, \Big)^{-1} \, ,
\end{equation}
\checked{rs}
where $a= \pm 1$ gives Fermi-Dirac or Bose-Einstein statistics, respectively, and $a=0$ gives Boltzmann statistics.  In the results section we will consider a Boltzmann distribution specifically, but in the interim we will not specify a particular form for $f_{\rm iso}$.

\subsection{Dynamical Variables}
\label{sect:bi1d}

At leading order in anisotropic hydrodynamics one assumes that \mbox{$\xi^{\mu\nu} = {\rm diag}(0,{\boldsymbol \xi})$} with ${\boldsymbol \xi} \equiv (\xi_x,\xi_y,\xi_z)$ which satisfy $\xi_x + \xi_y + \xi_z = 0$ due the tracelessness of $\xi^{\mu\nu}$.  In this case, expanding the argument of the square root appearing on the right-hand side of Eq.~(\ref{eq:genf}) in the local rest frame gives
\be
f(x,p) = f_{\rm iso}\!\left(\frac{1}{\lambda} \sqrt{p_\mu \Xi^{\mu\nu} p_\nu} \right) 
=  f_{\rm iso}\!\left(\frac{1}{\lambda}\sqrt{\sum_i \frac{p_i^2}{\alpha_i^2} + m^2}\right)  \, ,
\label{eq:fform}
\ee
\checked{rs}
where $i\in \{x,y,z\}$ and we have introduced the scale parameters
\be
\alpha_i \equiv (1 + \xi_i + \Phi)^{-1/2} \, .
\label{eq:alphadef}
\ee  
\checked{rs}
In the limit that $\xi_i = \Phi = 0$, $\alpha_i =1$ and one has $p_\mu \Xi^{\mu\nu} p_\nu = (p \cdot u)^2 = E^2$.  In practice, we will use the variables $\alpha_i$ as the dynamical parameters and then convert, after the fact, back to the $\xi_i$ and $\Phi$ when necessary.  In order for the quantity under the square root in (\ref{eq:fform}) to be positive for all possible momenta it suffices that
\ba
\xi_i \geq -1 - \Phi \, .
\ea
\checked{rs}
Since $\xi_z = - \xi_x - \xi_y$ this implies that
\ba
\xi_x + \xi_y \leq 1 + \Phi \, .
\ea
\checked{rs}
We note here that when one numerically solves the resulting dynamical equations, the constraints above are automatically satisfied by the dynamics.
We also mention that, using Eq.~(\ref{eq:alphadef}) and the tracelessness of the $\xi^{\mu\nu}$ tensor, one has
\be
\Phi = \frac{1}{3} \sum_i \alpha_i^{-2} - 1 \, .
\ee
\checked{s}

\subsection{Spheroidal form}
\label{app:varmap}

We note that, for a transversally homogeneous system, one can further simplify the distribution function by using $\xi_x = \xi_y \equiv \xi_\perp = -\xi_z/2$ and transforming to spheroidal form; however, it is frequently more convenient to perform this simplification at the end of the calculation.  In practice, therefore, we will use the general ellipsoidal form (\ref{eq:fform}) for the remainder of the paper.  That being said, when making connection to past works, the spheroidal form is still useful.  In the spheroidal case, one can parameterize the distribution function as 
\be
f(x,p) = f_{\rm iso}\!\left(\frac{1}{\Lambda}\sqrt{p_\perp^2 + (1+\xi) p_z^2 + (1+\tilde\Phi) m^2 } \right) .
\label{eq:rsf2}
\ee
\checked{rs}
Matching this to ellipsoidal form and using $\alpha_x = \alpha_y$ gives three equations
\ba
1 + \xi &=& \frac{\alpha_x^2}{\alpha_z^2} = \frac{1 + \xi_z + \Phi}{1 - \xi_z/2 + \Phi} \, ,
\nonumber \\
\Lambda &=& \alpha_x \lambda = \frac{\lambda}{\sqrt{1 - \xi_z/2 + \Phi}} \, ,
\nonumber \\
1+\tilde\Phi &=& \alpha_x^2 = \frac{1}{1 - \xi_z/2 + \Phi} \, ,
\label{eq:esrel1}
\ea 
\checked{rs}
from which one can obtain the following relations
\ba
\xi_z &=& \frac{2 \xi}{3(1 + \tilde\Phi)} \, , \nonumber \\
\Phi &=& \frac{\xi - 3\tilde\Phi}{3(1 + \tilde\Phi)} \, , \nonumber \\
\lambda &=& \frac{\Lambda}{\sqrt{1 + \tilde\Phi}} \, .
\label{eq:esrel2}
\ea
\checked{rs}
Note that in the limit $\Phi \rightarrow 0$ one obtains $\tilde\Phi = \xi/3$ and in
the limit $\tilde\Phi \rightarrow 0$ one obtains $\Phi = \xi/3$.

\section{Bulk variables}
\label{sect:bulkvars}

We now turn to the calculation of the bulk variables: number density, energy density, and the spacelike diagonal components of $T^{\mu\nu}$ (pressures).  These can be straightforwardly computed using the distribution function presented in the previous section.  For all quantities we write the invariant momentum-space integration measure as
\be
dP = \frac{d^3p}{(2\pi)^3} \frac{1}{E} \, .
\ee
\checked{rs}
In what follows, we will present general formulae and then specialize to the case of a Boltzmann distribution at the end.  With this in mind, we note that for a massive Boltzmann distribution, the isotropic equilibrium bulk variables are
\begin{eqnarray}
n_{\rm eq}(T,m) &=& 4 \pi \tilde{N} T^3 \, \hat{m}_{\rm eq}^2 K_2\left( \hat{m}_{\rm eq}\right) \,  , \label{eq:neq} \\
{\cal S}_{\rm eq}(T,m) &=& 4 \pi \tilde{N} T^3 \, \hat{m}_{\rm eq}^2 
 \Big[ 4 K_{2}\left( \hat{m}_{\rm eq} \right) + \hat{m}_{\rm eq} K_{1} \left( \hat{m}_{\rm eq}\right) \Big] \, ,
\label{eq:sigmaeq} \\
{\cal E}_{\rm eq}(T,m) &=& 4 \pi \tilde{N} T^4 \, \hat{m}_{\rm eq}^2
 \Big[ 3 K_{2}\left( \hat{m}_{\rm eq} \right) + \hat{m}_{\rm eq} K_{1} \left( \hat{m}_{\rm eq} \right) \Big] \, , 
\label{eq:epsiloneq} \\
{\cal P}_{\rm eq}(T,m) &=& n_{\rm eq}(T,m) \, T \, ,
\label{eq:Peq}
\end{eqnarray}
\checked{rs}
where $\hat{m}_{\rm eq} = m/T$ and $\tilde{N} = N_{\rm dof}/(2\pi)^3$ with $N_{\rm dof}$ being the number of degrees of freedom.

\subsection{Number Density}

Using (\ref{eq:fform}), the number density is 
\ba
n({\boldsymbol\xi},\Phi,m) &=& N_{\rm dof} \int dP E \, 
f_{\rm iso}\!\left(\frac{1}{\lambda}\sqrt{\sum_i \frac{p_i^2}{\alpha_i^2} + m^2}\right) 
\nonumber \\
&=& \alpha \, n_{\rm iso}(\lambda,m) \, ,
\label{eq:nequation}
\ea
\checked{rs}
where the sum over $i$ includes $i \in \{1,2,3\} = \{x,y,z\}$ and, for later convenience we have defined
\be
\alpha \equiv \prod_i \alpha_i  \, .
\ee
\checked{rs}
Note that if $f_{\rm iso}$ is a Boltzmann distribution, then $n_{\rm iso}(\lambda,m) = n_{\rm eq}(\lambda,m)$ 
with $n_{\rm eq}(\lambda,m)$ specified by Eq.~(\ref{eq:neq}).  

\subsection{Energy Density}

The energy density is given by 
\ba
{\cal E} &=& N_{\rm dof} \int dP E^2 \, f_{\rm iso}\!\left(\frac{1}{\lambda}\sqrt{\sum_i \frac{p_i^2}{\alpha_i^2} + m^2}\right)
\nonumber  \\
&=& \tilde{N} \int d^3p \; 
    \sqrt{{\bf p}^2 + m^2}
    \, f_{\rm iso}\!\left(\frac{1}{\lambda}\sqrt{\sum_i \frac{p_i^2}{\alpha_i^2} + m^2}\right) . \;\;\;
\label{eq:edensint}
\ea
\checked{rs}
Changing variables to $\hat{p}_i = p_i/(\lambda \alpha_i)$ and transforming to spherical coordinates one obtains
\be
{\cal E} = {\cal H}_3({\boldsymbol\xi},\Phi,\hat{m}) \, \lambda^4 \, ,
\label{eq:edens}
\ee
\checked{rs}
where the ${\cal H}_3$ function appearing above is defined in Eq.~(\ref{eq:h3gen}) and $\hat{m} \equiv m/\lambda$.  

For a transversally homogeneous system one has $\alpha_x = \alpha_y$.  In this case, one has alternatively
\be
{\cal E} = \tilde{\cal H}_3({\boldsymbol \xi},\Phi,\hat{m}) \, \lambda^4 \, ,
\label{eq:edenst}
\ee
\checked{rs}
with $\tilde{\cal H}_3$ defined in Eq.~(\ref{eq:h3tilde}).  

\subsection{Transverse Pressure}

The transverse pressure is given by 
\ba
{\cal P}_T &=& \frac{N_{\rm dof}}{2} \int dP \, (p_x^2+p_y^2) \, f_{\rm iso}\!\left(\frac{1}{\lambda}\sqrt{\sum_i \frac{p_i^2}{\alpha_i^2} + m^2}\right)
\nonumber  \\
&=& \frac{\tilde{N}}{2} \int d^3p \; 
    \frac{p_x^2+p_y^2}{\sqrt{{\bf p}^2 + m^2}}
    \, f_{\rm iso}\!\left(\frac{1}{\lambda}\sqrt{\sum_i \frac{p_i^2}{\alpha_i^2} + m^2}\right) .
\label{eq:ptint}
\ea
\checked{rs}
Again changing variables to $\hat{p}_i = p_i/(\lambda \alpha_i)$ and transforming to spherical coordinates one obtains
\be
{\cal P}_T = {\cal H}_{3T}({\boldsymbol\xi},\Phi,\hat{m}) \, \lambda^4 \, ,
\ee
\checked{rs}
where the ${\cal H}_{3T}$ function appearing above is defined in Eq.~(\ref{eq:h3tgen}).  When the system is transversally homogeneous one obtains
\be
{\cal P}_T = \tilde{\cal H}_{3T}({\boldsymbol \xi},\Phi,\hat{m}) \, \lambda^4 \, ,
\ee
\checked{rs}
where the $\tilde{\cal H}_{3T}$ function appearing above is defined in Eq.~(\ref{eq:h3ttilde}).

\subsection{Longitudinal Pressure}

The longitudinal pressure is given by 
\ba
{\cal P}_L &=& N_{\rm dof} \int dP \, p_z^2 \, f_{\rm iso}\!\left(\frac{1}{\lambda}\sqrt{\sum_i \frac{p_i^2}{\alpha_i^2} + m^2}\right)
\nonumber  \\
&=& \tilde{N} \int d^3p \; 
    \frac{p_z^2}{\sqrt{{\bf p}^2 + m^2}}
    \, f_{\rm iso}\!\left(\frac{1}{\lambda}\sqrt{\sum_i \frac{p_i^2}{\alpha_i^2} + m^2}\right) .
\label{eq:plint}
\ea
\checked{rs}
Again changing variables to $\hat{p}_i = p_i/(\lambda \alpha_i)$ and transforming to spherical coordinates one obtains
\be
{\cal P}_L = {\cal H}_{3L}({\boldsymbol\xi},\Phi,\hat{m}) \, \lambda^4 \, ,
\label{eq:pl}
\ee
\checked{rs}
where the ${\cal H}_{3L}$ function appearing above is defined in Eq.~(\ref{eq:h3lgen}).  When the system is transversally homogeneous one obtains
\be
{\cal P}_L = \tilde{\cal H}_{3L}({\boldsymbol \xi},\Phi,\hat{m}) \, \lambda^4 \, ,
\label{eq:plt}
\ee
\checked{rs}
where the $\tilde{\cal H}_{3L}$ function appearing above is defined in Eq.~(\ref{eq:h3ltilde}).

\section{Moments of the Boltzmann equation}
\label{sect:beqmoments}

To obtain the necessary equations of motion, we take moments of the Boltzmann equation in the relaxation time approximation
\be
p^\mu \partial_\mu f = \frac{1}{\tau_{\rm eq}} \left( f_{\rm iso} - f \right) ,
\label{eq:beqrta}
\ee
where $f_{\rm iso}$ is the late-time isotropic equilibrium fixed point and $\tau_{\rm eq}$ is the relaxation time which herein we assume to be constant.  In the subsections below, we compute the zeroth, first, and second moments of Eq.~(\ref{eq:beqrta}) for a boost-invariant cylindrically symmetric system.  At the end of each subsection, we specify the simpler equation that results if the system is, in addition, transversally homogeneous.  Finally, for each moment we simplify to the case that the underlying isotropic distribution function is given by a Boltzmann distribution.

\subsection{Zeroth Moment}
\label{sect:0mom}

Computing the zeroth moment of Eq.~(\ref{eq:beqrta}) gives
\be
D n + n \theta = \frac{1}{\tau_{\rm eq}} ( n_{\rm iso} - n ) \, . 
\label{eq:zeromom1}
\ee
\checked{rs}
For one-dimensional transversally homogeneous expansion one has $D = \partial_\tau$ and $\theta = 1/\tau$ giving
\be
\partial_\tau n +  \frac{n}{\tau} = \frac{1}{\tau_{\rm eq}} ( n_{\rm iso} - n ) \, ,
\ee
\checked{rs}
which upon using (\ref{eq:nequation}) becomes
\be
\partial_\tau \log \alpha_x^2 \alpha_z
+ (\partial_\lambda \log n_{\rm iso} ) \partial_\tau \lambda + \frac{1}{\tau}
=  \frac{1}{\tau_{\rm eq}} \left[ \frac{1}{\alpha_x^2 \alpha_z} \frac{n_{\rm iso}(T,m)}{n_{\rm iso}(\lambda,m)} - 1 \right] \, .
\ee
\checked{rs}
Finally, specializing to the case that $f_{\rm iso}$ is a Boltzmann distribution one obtains
\be
\partial_\tau \log \alpha_x^2 \alpha_z
+ \left[ 3 + \hat{m} \frac{K_1(\hat{m})}{K_2(\hat{m})} \right] \, \partial_\tau \log \lambda + \frac{1}{\tau}
=  \frac{1}{\tau_{\rm eq}} \left[ \frac{1}{\alpha_x^2 \alpha_z} \frac{T}{\lambda}\frac{K_2(\hat{m}_{\rm eq})}{K_2(\hat{m})} - 1 \right] .
\label{eq:final0mom}
\ee
\checked{rs}

\subsection{First Moment}
\label{sect:1mom}

The first moment of Eq.~(\ref{eq:beqrta}) gives energy-momentum conservation
\be
\partial_\mu T^{\mu\nu} = 0 \, .
\label{eq:firstmom}
\ee
\checked{rs}
The vanishing of the right hand side in (\ref{eq:firstmom}) results in the dynamical Landau matching condition
\be
u_\mu T^{\mu\nu} = u_\mu T^{\mu\nu}_{\rm eq} \, .
\label{eq:dynlandau}
\ee
\checked{rs}
Here $T^{\mu \nu}_{\rm eq}$ is the equilibrium energy-momentum tensor  
\ba
T^{\mu \nu}_{\rm eq} = \left( {\cal E}_{\rm eq}
+ {\cal P}_{\rm eq} \right)  u^\mu u^\nu
- {\cal P}_{\rm eq} g^{\mu\nu}  \, ,
\label{eq:TEQ}
\ea
\checked{rs}
where ${\cal E}_{\rm eq}$ and ${\cal P}_{\rm eq}$ are given by Eqs.~(\ref{eq:epsiloneq}) and (\ref{eq:Peq}), respectively, in the case of Boltzmann statistics.  
We will return to the issue of dynamical Landau matching shortly and present the nonlinear
equation which must be solved in order to enforce this constraint.

For a boost-invariant and cylindrically symmetric system the energy-momentum tensor $T^{\mu\nu}$ has the general structure
\be
T^{\mu \nu} = {\cal E} u^\mu u^\nu
+ {\cal P}_x x^\mu x^\nu
+ {\cal P}_y y^\mu y^\nu
+ {\cal P}_z z^\mu z^\nu \, .
\label{eq:TAHg}
\ee
\checked{rs}
The resulting dynamical equations in this case are~\cite{Tinti:2013vba}
\ba
D {\cal E} + {\cal E} \theta &=& \sum_i {\cal P}_i \theta_i \, , \\
D_x {\cal P}_x + {\cal P}_x (\partial_\mu x^\mu) &=& {\cal E} (x_\mu D u^\mu) 
+ {\cal P}_y (x_\mu D_y y^\mu) + {\cal P}_z (x_\mu D_z z^\mu) \, ,
\ea
\checked{rs}
where $D = u^\mu \partial_\mu$, $\theta = \partial_\mu u^\mu$, and $D_i = X_i^\mu \partial_\mu$.

For a transversally homogeneous system one can take $\alpha_x = \alpha_y$ and the energy-momentum tensor $T^{\mu\nu}$ has a somewhat simpler structure
\be
T^{\mu \nu} = \left( {\cal E}
+ {\cal P}_T \right)  u^\mu u^\nu
- {\cal P}_T g^{\mu\nu}
+\left( {\cal P}_L - {\cal P}_T \right) z^\mu z^\nu \, .
\label{eq:TAH}
\ee
\checked{rs}
Further assuming boost-invariance, the equations of motion reduce to
\be
\partial_\tau {\cal E} = - \frac{{\cal E} + {\cal P}_L}{\tau} \, .
\label{eq:firstmom1D}
\ee
\checked{rs}
Using Eqs.~(\ref{eq:edenst}) and (\ref{eq:plt}) this becomes explicitly
\ba
\partial_\tau \left[ \tilde{\cal H}_3({\boldsymbol \xi},\Phi,\hat{m}) \, \lambda^4 \right]
= -\frac{\lambda^4}{\tau} \left[ \tilde{\cal H}_3({\boldsymbol \xi},\Phi,\hat{m})  
+ \tilde{\cal H}_{3L}({\boldsymbol \xi},\Phi,\hat{m}) \right] .
\ea
\checked{rs}
Expanding the left hand side one obtains
\be
\partial_\tau \left( \tilde{\cal H}_3 \, \lambda^4 \right)  
= \lambda^4 \left[
(\partial_{\alpha_x} \tilde{\cal H}_3) \, \partial_\tau \alpha_x
+ 
(\partial_{\alpha_z} \tilde{\cal H}_3) \, \partial_\tau \alpha_z
+ 
(\partial_{\lambda} \tilde{\cal H}_3) \, \partial_\tau \lambda 
+
4 \tilde{\cal H}_3 \partial_\tau \log\lambda \right]  .
\ee
\checked{rs}
To evaluate the necessary derivatives one can use the following identities
\ba
\frac{\partial {\cal H}_2(y,z)}{\partial y} &=& \frac{1}{y} \Bigl[ {\cal H}_2(y,z) + {\cal H}_{2L}(y,z) \Bigr] ,
\nonumber \\
\frac{\partial {\cal H}_2(y,z)}{\partial z} &=& \frac{1}{z} \Bigl[ {\cal H}_2(y,z) - {\cal H}_{2L}(y,z) - {\cal H}_{2T}(y,z) \Bigr] ,
\ea
\checked{rs}
which can be used to show that
\ba
\frac{\partial \tilde{\cal H}_3}{\partial \alpha_x} &=& \frac{2}{\alpha_x} \Bigl( \tilde{\cal H}_3 + \tilde{\cal H}_{3T} \Bigr) \equiv \frac{2}{\alpha_x} \tilde\Omega_T \, ,
\nonumber \\
\frac{\partial \tilde{\cal H}_3}{\partial \alpha_z} &=& \frac{1}{\alpha_z} \Bigl( \tilde{\cal H}_3 + \tilde{\cal H}_{3L} \Bigr) \equiv \frac{1}{\alpha_z} \tilde\Omega_L \, ,
\nonumber \\
\frac{\partial \tilde{\cal H}_3}{\partial \hat{m}} &=& \frac{1}{\hat{m}} \Bigl( \tilde{\cal H}_3 -\tilde{\cal H}_{3L} - 2 \tilde{\cal H}_{3T} - \tilde{\cal H}_{3m} \Bigr) \equiv \frac{1}{\hat{m}} \tilde\Omega_m \, ,
\label{eq:mintdef}
\ea
\checked{rs}
where the $\tilde{\cal H}$ functions above are understood to be evaluated at $({\boldsymbol \xi},\Phi,\hat{m})$ and $\tilde{\cal H}_{3m}$ is defined in Eq.~(\ref{eq:h3mtilde}).  The final result for the first moment equation is quite compact when written in terms of the special functions introduced above
\be
\left( 4 \tilde{\cal H}_3 - \tilde\Omega_m \right) \partial_\tau \log\lambda
+ \tilde\Omega_T \partial_\tau \log\alpha_x^2
+ \tilde\Omega_L \partial_\tau \log\alpha_z = 
-\frac{1}{\tau} \tilde\Omega_L \, .
\label{eq:final1mom}
\ee
\checked{rs}

\subsection{Second Moment}
\label{sect:2mom}

Computing the second moment of Eq.~(\ref{eq:beqrta}) one finds
\be
\partial_\lambda \Theta^{\lambda\mu\nu} = \frac{1}{\tau_{\rm eq}} \left(u_\lambda\Theta_{\rm eq}^{\lambda\mu\nu} - u_\lambda\Theta^{\lambda\mu\nu}\right),
\label{eq:tmom}
\ee
\checked{rs}
where 
\ba
\Theta^{\mu\nu\lambda} &= N_{\rm dof} \int \! dP \; p^\mu p^\nu p^\lambda  f \, ,
\label{eq:Thetas1}
\\
\Theta^{\mu\nu\lambda}_{\rm eq} &= N_{\rm dof} \int \! dP \; p^\mu p^\nu p^\lambda f_{\rm eq} \, .
\label{eq:Thetas2}
\ea
\checked{rs}
For distribution functions of the form specified in Eq.~(\ref{eq:fform}), the only non-vanishing terms in (\ref{eq:Thetas1}) and (\ref{eq:Thetas2}) are those that have an even number of each spatial index. In covariant form they read
\ba
\Theta &=& \Theta_u \left[ u\otimes u \otimes u\right] 
\nonumber \\
&& \,+\, \Theta_x \left[ u\otimes x \otimes x +x\otimes u \otimes x + x\otimes x \otimes u\right] 
\nonumber \\ 
&& \,+\,  \Theta_y  \left[ u\otimes y \otimes y +y\otimes u \otimes y + y\otimes y \otimes u\right]
\nonumber \\
&& \,+\, \Theta_z \left[ u\otimes z \otimes z +z\otimes u \otimes z + z\otimes z \otimes u\right] .
 \label{eq:Theta}
\ea
\checked{rs}
The equilibrium tensor has the same decomposition but, due to the rotational invariance of the local equilibrium state, one has $\Theta_{\rm iso} = \Theta_x = \Theta_y = \Theta_z$.

Evaluating the necessary integrals using the distribution function (\ref{eq:fform}), one finds
\be
\Theta_u = \left(\sum_i \alpha_i^2\right) \alpha \, \Theta_{\rm iso}(\lambda,m) + \alpha m^2 n_{\rm iso}(\lambda,m) \, ,
\ee
\checked{rs}
and
\be
\Theta_i = \alpha \, \alpha_i^2 \, \Theta_{\rm iso}(\lambda,m) \, ,
\ee
\checked{rs}
with
\be
\Theta_{\rm iso}(\lambda,m) \equiv \frac{4\pi\tilde{N}\lambda^5}{3}  \int_0^\infty \, d\hat{p} \, \hat{p}^4  f_{\rm iso}\!\left(\!\sqrt{\hat{p}^2 + \hat{m}^2}\,\right) ,
\ee
\checked{rs}
which for Boltzmann statistics becomes
\be
\Theta_{\rm iso}(\lambda,m) = \Theta_{\rm eq} =  4 \pi {\tilde N} \lambda^5 \hat{m}^3 K_3(\hat{m}) \, .
\ee
\checked{rs}
Note that, in general, one has
\be
\Theta_u - \sum_i \Theta_i = \alpha m^2 n_{\rm iso}(\lambda,m) \, ,
\label{eq:thetauID}
\ee
\checked{rs}
and in the limit $\lim_{m \rightarrow 0} \Theta_u = \sum_i \Theta_i$.

\subsubsection{Dynamical Equations}

We begin by considering the left hand side of (\ref{eq:tmom}).  Using the tensor decomposition (\ref{eq:Theta}) one obtains
\ba
\partial_\lambda \Theta^{\lambda \mu \nu} &=&
u^\mu u^\nu D \Theta_u
+ \Theta_u \left[ u^\mu u^\nu \theta + 2 u^{(\nu} D u^{\mu)} \right]
\nonumber \\ && 
+ x^\mu x^\nu D \Theta_x + 2 u^{(\mu} x^{\nu)} D_x \Theta_x
\nonumber \\ &&
+ \Theta_x \left[ x^\mu x^\nu \theta + 2 x^{(\nu} D x^{\mu)} \right] 
\nonumber \\ &&
+ 2 \Theta_x \left[ u^{(\mu} x^{\nu)} \partial_\alpha x^\alpha + x^{(\nu} D_x u^{\mu)}  + u^{(\mu} D_x x^{\nu)} \right] 
\nonumber \\ &&
+ ( x \rightarrow y) + (x \rightarrow z) \, ,
\label{eq:dTheta}
\ea
\checked{rs}
where, as before, $D = u^\mu \partial_\mu$, $\theta = \partial_\mu u^\mu$, and $D_i = X_i^\mu \partial_\mu$.

\subsubsection{$uu$ projection}

Projecting the left hand side of (\ref{eq:dTheta}) with $u_\mu u_\nu$ we obtain
\be
u_\mu u_\nu \partial_\lambda \Theta^{\lambda \mu \nu} =
D \Theta_u + \theta \Theta_u - 2 \sum_i \Theta_i \theta_i \, ,
\ee
\checked{rs}
where we have used $u_\mu D u^\mu = D(u_\mu u^\mu)/2 = 0$ and defined $\theta_i \equiv -u_\mu D_i X_i^\mu$.  Setting this equal to the right hand side of (\ref{eq:tmom}) we obtain
\ba
D \Theta_u + \theta \Theta_u  - 2 \sum_i \Theta_i \theta_i
= \frac{1}{\tau_{\rm eq}} ( \Theta_{u,\rm eq} - \Theta_u ) \, .
\ea
\checked{rs}

\subsubsection{$ux$ projection}

Projecting the left hand side of of (\ref{eq:dTheta}) with $u_\mu x_\nu$ we obtain
\ba
u_\mu x_\nu \partial_\lambda \Theta^{\lambda \mu \nu} &=&
\Theta_u ( x_\nu D u^\nu ) - D_x \Theta_x - \Theta_x ( u_\mu D x^\mu)  - \Theta_x (\partial_\mu x^\mu)
\nonumber \\ && \hspace{1cm}
+ \Theta_y (x_\nu D_y y^\nu) + \Theta_z (x_\nu D_z z^\nu) \, ,
\ea
\checked{rs}
which gives 
\be
(\Theta_u + 2 \Theta_x) D \theta_\perp
+ D_x \Theta_x = \frac{\cosh\theta_\perp}{r} (\Theta_y - \Theta_x) + 
\frac{\sinh\theta_\perp}{\tau} (\Theta_z - \Theta_x) \, .
\ee
\checked{rs}

\subsubsection{$xx$, $yy$, and $zz$ projections}

Projecting the left hand side of (\ref{eq:dTheta}) with $x_\mu x_\nu$ we obtain
\be
x_\mu x_\nu \partial_\lambda \Theta^{\lambda \mu \nu} =
D \Theta_x + \Theta_x \theta - 2 \Theta_x ( x_\mu D_x u^\mu ) \, ,
\ee
\checked{rs}
so that the $xx$ equation becomes
\be
D \Theta_x + \Theta_x (\theta - 2 \theta_x) 
= \frac{1}{\tau_{\rm eq}} ( \Theta_{\rm eq} - \Theta_x ) \, .
\ee
\checked{rs}
Likewise
\be
y_\mu y_\nu \partial_\lambda \Theta^{\lambda \mu \nu} =
D \Theta_y + \Theta_y (\theta - 2 \theta_y) = \frac{1}{\tau_{\rm eq}} ( \Theta_{\rm eq} - \Theta_y ) \, ,
\ee
\checked{rs}
and
\be
z_\mu z_\nu \partial_\lambda \Theta^{\lambda \mu \nu} =
D \Theta_z + \Theta_z (\theta - 2 \theta_z) = \frac{1}{\tau_{\rm eq}} ( \Theta_{\rm eq} - \Theta_z ) \, .
\ee
\checked{rs}

\subsubsection{$uy$, $uz$, $xy$, $xz$, and $yz$ projections}

Projecting the left hand side of (\ref{eq:dTheta}) with $u_\mu y_\nu$ we obtain
\be
D_y \Theta_y = 0 \, ,
\ee
which, upon using Eq.~(\ref{eq:derivs1}), gives $\partial_\phi \Theta_y = 0$ which is trivially satisfied due to the assumed cylindrical symmetry.  Likewise, the $uz$-projection gives
\be
D_z \Theta_z = 0 \, ,
\ee
which, upon using Eq.~(\ref{eq:derivs1}), gives $\partial_{\eta_\parallel} \Theta_z = 0$ which is trivially satisfied due to boost invariance.  The remaining off-diagonal spacelike projections ($xy$, $xz$, and $yz$) can also be shown to be 
trivially satisfied.

\subsubsection{Final second moment equations}

Summarizing, for a boost-invariant cylindrically symmetric system one obtains the following (non-trivial) dynamical equations from the second moment~\cite{Tinti:2013vba}\footnote{There appears to be a typo in the analog of (\ref{eq:thetar}) in Ref.~\cite{Tinti:2013vba}.}
\ba
D \Theta_u + \theta \Theta_u  - 2 \sum_i \Theta_i \theta_i
&=& \frac{1}{\tau_{\rm eq}} ( \Theta_{u,\rm eq} - \Theta_u ) \, ,
\label{eq:thetau}
\\
(\Theta_u + 2 \Theta_x) D \theta_\perp
+ D_x \Theta_x &=& \frac{\cosh\theta_\perp}{r} (\Theta_y - \Theta_x) + 
\frac{\sinh\theta_\perp}{\tau} (\Theta_z - \Theta_x) \, ,
\label{eq:thetar}
\\
D \Theta_i + \Theta_i (\theta - 2 \theta_i) 
&=& \frac{1}{\tau_{\rm eq}} ( \Theta_{\rm eq} - \Theta_i ) \, ,
\label{eq:thetai}
\ea
\checked{rs}
where $i \in \{x,y,z\}$.

For the case of boost-invariant cylindrically symmetric expansion, there are five equations coming from the second moment which, when combined with the zeroth and first moment equations, gives eight equations for five unknowns ($\xi_x$, $\xi_z$, $\Phi$, $\lambda$, and $\theta_\perp$).  However, two of these equations are related.  To see this, consider the $\Theta_u$ equation (\ref{eq:thetau}).  Using Eq.~(\ref{eq:thetauID}) in (\ref{eq:thetau}) and then substituting the equations of motion for $\Theta_i$ (\ref{eq:thetai}) one finds 
\be
m^2 \biggl[ D n + n \theta \biggr] = m^2 \left[ \frac{1}{\tau_{\rm eq}} ( n_{\rm iso} - n ) \right] . 
\ee
\checked{rs}
For massless systems this is satisfied trivially, however, if $m$ is finite, one has
\be
D n + n \theta  =  \frac{1}{\tau_{\rm eq}} ( n_{\rm iso} - n )  \, ,
\ee
\checked{rs}
which is precisely the zeroth moment equation obtained previously (\ref{eq:zeromom1}).  Since the second moment equation for $\Theta_u$ is identical to the zeroth moment equation, this leaves us with seven equations for five unknowns.  As demonstrated by Tinti and Florkowski \cite{Tinti:2013vba}, the three equations for $\Theta_i$ (\ref{eq:thetai}) can be reduced to two equations since the third is guaranteed if the other two are satisfied.  This leaves us with six equations for five unknowns.  To proceed, one can follow the suggestion of Tinti and Florkowski, which is to disregard the $ux$-projection equation (\ref{eq:thetar}).  If one follows this prescription, we then have the same number of equations as unknowns, namely five.  We will return to this issue in the conclusions.

The situation is somewhat simpler for a 0+1d massive gas and the system of equations closes without having to make such choices.  Using the four vectors specified in Sec.~\ref{sect:basis} we can now specialize to the case of boost-invariant transversally homogeneous expansion for which one has: 
$\theta_\perp=0$, $\theta=1/\tau$, $\theta_x=0$, $\theta_y=0$, and $\theta_z=-1/\tau$. In this case Eq.~(\ref{eq:thetar}) is trivially satisfied, eliminating it as a dynamical equation, and Eq.~(\ref{eq:thetai}) can be evaluated for any $i$.  For example, assuming Boltzmann statistics, the ``raw'' $\Theta_x$ equation is\footnote{The $\Theta_y$ equation is the same for a transversally homogeneous system by symmetry.}
\be
\left[ 5 + \hat{m} \frac{K_2(\hat{m})}{K_3(\hat{m})} \right] \partial_\tau \log\lambda + 4 \partial_\tau \log\alpha_x + \partial_\tau\log\alpha_z + \frac{1}{\tau} = \frac{1}{\tau_{\rm eq}} \left( \frac{\Theta_{\rm eq}}{\Theta_x}-1\right) ,
\ee
\checked{s}
with a similar result for $\Theta_z$.
To simplify things further, we follow \cite{Tinti:2013vba} and subtract one third of the sum of the $\Theta_i$ equations from each of the $\Theta_i$ equations.\footnote{Following Ref.~\cite{Tinti:2013vba} we also discard the equation implied by the sum of the $\Theta_i$ equations.}  When this is done, the equations obtained with $i = x,y,z$ can be shown to be equivalent.  As a consequence, the final second moment equation necessary to describe the 0+1d evolution is
\be
\partial_\tau \log\left( \frac{\alpha_x}{\alpha_z} \right) - \frac{1}{\tau} + \frac{3}{4\tau_{\rm eq}} \frac{\xi_z}{\alpha_x^2\alpha_z} \left( \frac{T}{\lambda} \right)^2 \frac{K_3(\hat{m}_{\rm eq})}{K_3(\hat{m})} = 0 \, .
\label{eq:final2mom}
\ee
\checked{s}
where $\xi_z = \frac{2}{3}(\alpha_z^{-2} - \alpha_x^{-2})$.

\subsection{0+1d Dynamical Equations}

Our final set of three dynamical equations which describe the evolution of a massive 0+1d system including the effect of bulk viscous pressure are given by Eqs.~(\ref{eq:final0mom}), (\ref{eq:final1mom}), and (\ref{eq:final2mom}).  We collect them together here
\ba
&& \partial_\tau \log \alpha_x^2 \alpha_z
+ \left[ 3 + \hat{m} \frac{K_1(\hat{m})}{K_2(\hat{m})} \right] \, \partial_\tau \log \lambda + \frac{1}{\tau}
=  \frac{1}{\tau_{\rm eq}} \left[ \frac{1}{\alpha_x^2 \alpha_z} \frac{T}{\lambda}\frac{K_2(\hat{m}_{\rm eq})}{K_2(\hat{m})} - 1 \right] \, ,
\label{eq:final0m}
\\
&&
\left( 4 \tilde{\cal H}_3 - \tilde\Omega_m \right) \partial_\tau \log\lambda
+ \tilde\Omega_T \partial_\tau \log\alpha_x^2
+ \tilde\Omega_L \partial_\tau \log\alpha_z =
-\frac{1}{\tau} \tilde\Omega_L \, ,
\label{eq:final1m}
\\
&&
 \partial_\tau \log\left( \frac{\alpha_x}{\alpha_z} \right) - \frac{1}{\tau} + \frac{3}{4\tau_{\rm eq}} \frac{\xi_z}{\alpha_x^2\alpha_z} \left( \frac{T}{\lambda} \right)^2 \frac{K_3(\hat{m}_{\rm eq})}{K_3(\hat{m})} = 0 \, ,
\label{eq:final2m}
\ea
\checked{last eq tweaked - recheck}
where we have specialized to the case that the underlying isotropic distribution function is a Boltzmann distribution.
These three equations can be used to evolve $\xi_z$, $\Phi$, and $\lambda$ in proper-time (or alternatively $\alpha_x$, $\alpha_z$, and $\lambda$).  
The catch, however, is that these equations involve the effective temperature $T$.  For this purpose, one can use  Eq.~(\ref{eq:dynlandau}) to determine $T$ in terms of the underlying microscopic parameters in the distribution function at any moment in time.  This results in the constraint equation
\be
\tilde{\cal H}_3 \lambda^4 = 4 \pi \tilde{N} T^4 \hat{m}_{\rm eq}^2 
 \Big[ 3 K_{2}\left( \hat{m}_{\rm eq} \right) + \hat{m}_{\rm eq} K_{1} \left( \hat{m}_{\rm eq} \right) \Big] .
\label{eq:dlm}
\ee
\checked{s}
%

\begin{figure}[t]
\centerline{\includegraphics[angle=0,width=0.7\textwidth]{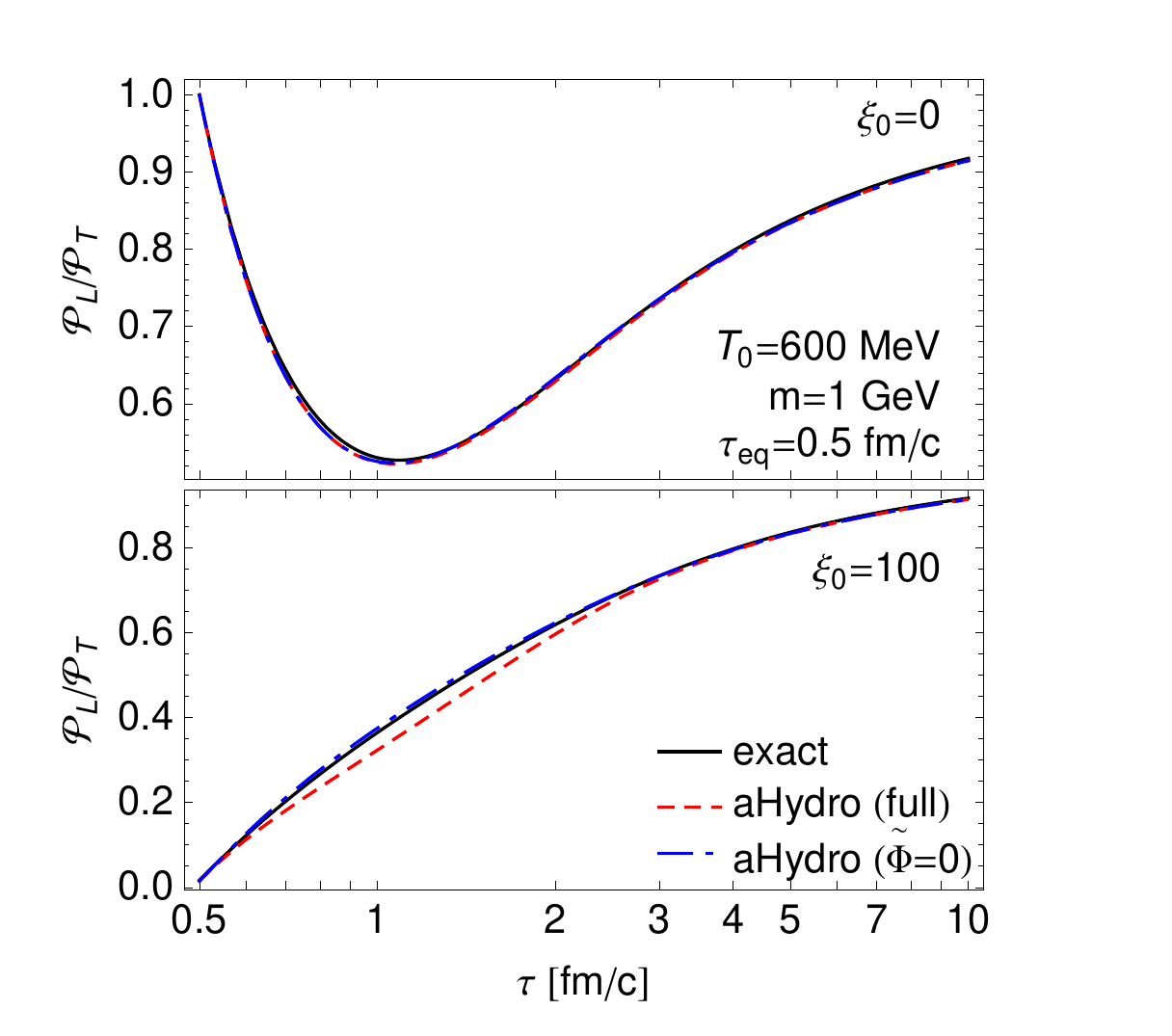}}
\caption{(Color online) Proper-time evolution of ${\cal P}_L/{\cal P}_T$.  The three lines correspond to the exact solution of the Boltzmann equation \cite{Florkowski:2014sfa} (black solid line), the full aHydro equations including the bulk degree of freedom (red dashed line), and the aHydro equations with the ellipsoidal bulk degree of freedom set to zero (blue dot-dashed line).  For both panels we used $m=$ 1 GeV, $\tau_0$ = 0.5 fm/c, $\tau_{\rm eq}$ = 0.5 fm/c, and $T_0$ = 600 MeV.  In the top panel we fixed the initial spheroidal anisotropy parameter $\xi_0=0$ and in the bottom panel we chose $\xi_0 = 100$. 
}
\label{fig:PL2PT_1000}
\end{figure}

We note that instead of using a root solver to enforce (\ref{eq:dlm}), it is possible to transform (\ref{eq:dlm}) into a differential equation by taking a derivative with respect to $\tau$ on the left and right hand sides~\cite{Florkowski:2012as}.  Since the left hand side then simply becomes the left hand side of the energy conservation equation, we can use (\ref{eq:final1m}) to simplify the result giving 
\be
\partial_\tau \left\{ 4 \pi \tilde{N} T^4 \hat{m}_{\rm eq}^2 
 \Big[ 3 K_{2}\left( \hat{m}_{\rm eq} \right) + \hat{m}_{\rm eq} K_{1} \left( \hat{m}_{\rm eq} \right) \Big] \right\} 
= -\frac{1}{\tau} \tilde\Omega_L \lambda^4 ,
\ee
\checked{s}
which simplifies to
\be
\partial_\tau \log T = - \frac{1}{\tau} \frac{\lambda^4}{T^4} \frac{\tilde\Omega_L}{\tilde\Omega_{\rm eq}} \, ,
\label{eq:coneq}
\ee
\checked{s}
where
\be
\tilde\Omega_{\rm eq} \equiv 4 \pi \tilde{N} \hat{m}_{\rm eq}^2 \left[ 12 K_2(\hat{m}_{\rm eq}) + 5 \hat{m}_{\rm eq} K_1(\hat{m}_{\rm eq}) + \hat{m}_{\rm eq}^2 K_0(\hat{m}_{\rm eq}) \right] .
\ee
\checked{s}
If one uses this method, one needs to ensure that Eq.~(\ref{eq:dlm}) is satisfied at $\tau=\tau_0$ and then one can evolve the constraint equation (\ref{eq:coneq}) along with Eqs.~(\ref{eq:final0m}), (\ref{eq:final1m}), and (\ref{eq:final2m}) as an additional dynamical equation.  We will use both methods to check our numerical results, but will primarily use the root-finding method since, in practice, it is slightly more numerically efficient for the case at hand.

\section{Numerical results}
\label{sect:results}

\begin{figure}[t]
\centerline{\includegraphics[angle=0,width=0.7\textwidth]{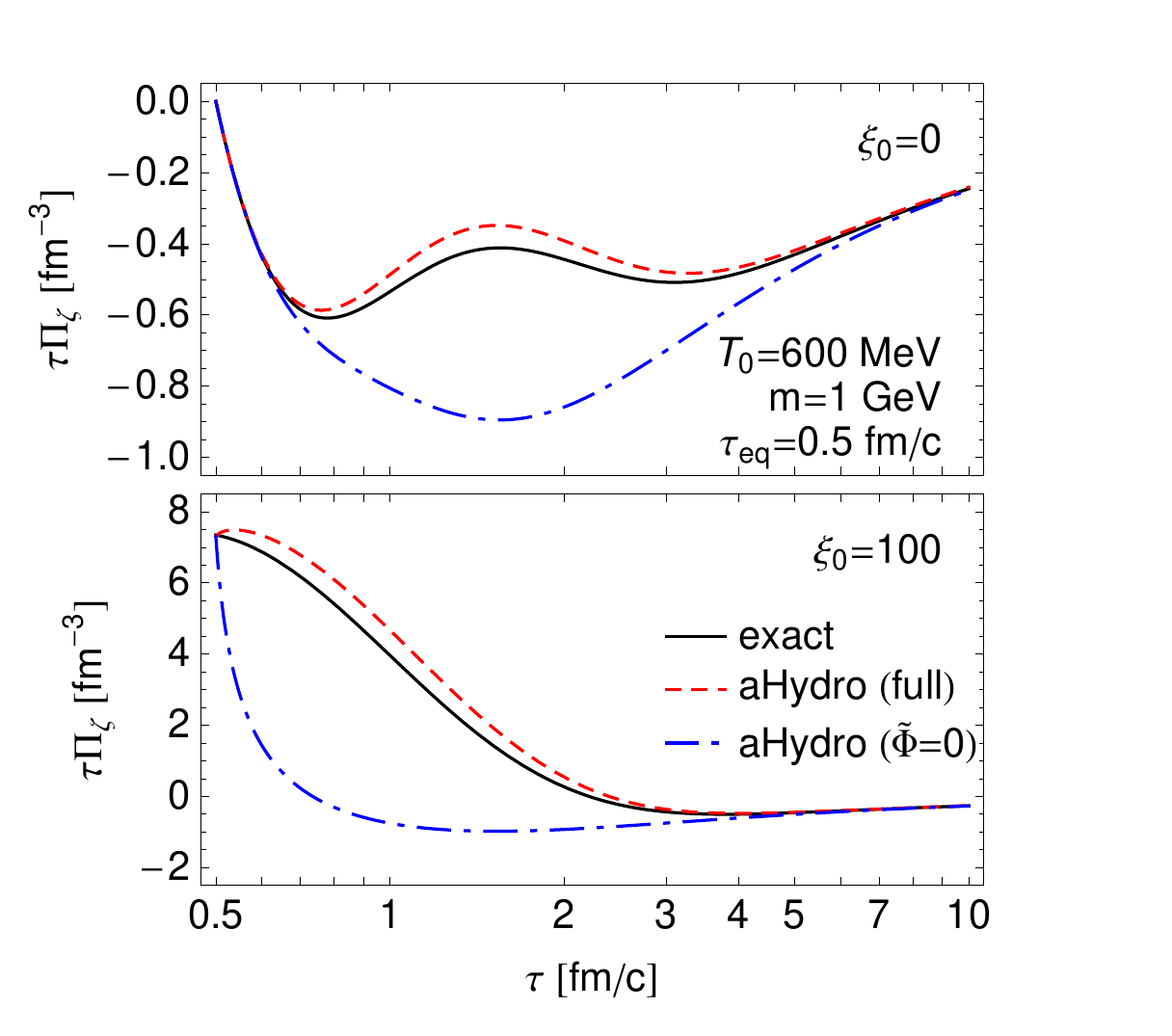}}
\caption{(Color online) Proper-time evolution of the bulk pressure.  Parameters and descriptions are the same as in Fig.~\ref{fig:PL2PT_1000}.}
\label{fig:bulk_1000}
\end{figure}

We now compare the evolution predicted by Eqs.~(\ref{eq:final0m}), (\ref{eq:final1m}), and (\ref{eq:final2m}) with the exact solution of the massive Boltzmann equation recently obtained in Ref.~\cite{Florkowski:2014sfa}.  Instead of evolving the anisotropy parameter $\xi_z$ and bulk parameter $\Phi$ we instead evolve $\alpha_x$ and $\alpha_z$ numerically.  We fix the initial conditions for $\alpha_x^0$, $\alpha_z^0$, and $T_0=$ 600 MeV at $\tau_0 = 0.5$ fm/c and fix $\lambda_0$ using Eq.~(\ref{eq:dlm}).  We then use Eqs.~(\ref{eq:final0m}), (\ref{eq:final1m}), and (\ref{eq:final2m}) to evolve $\alpha_{x}$, $\alpha_{y}$, and $\lambda$.  At each step of the numerical integration we use Eq.~(\ref{eq:dlm}) to self-consistently determine the effective temperature $T$ which appears in the equations of motion or, alternatively, evolve the temperature using Eq.~(\ref{eq:coneq}).

In Figs.~\ref{fig:PL2PT_1000} and \ref{fig:bulk_1000} we plot the proper-time evolution of ${\cal P}_L/{\cal P}_T$ and the bulk pressure $\Pi_\zeta$, respectively.  The bulk pressure is computed via
\be
\Pi_\zeta(\tau) = \frac{1}{3}
\left[{\cal P}_L(\tau) + 2 {\cal P}_T(\tau)
- 3 {\cal P}_{\rm eq}(\tau) \right] ,
\label{eq:PIkz}
\ee
\checked{s}
where ${\cal P}_{\rm eq}$ is the equilibrium pressure evaluated at the effective temperature $T(\tau)$.  In both figures the three lines correspond to the exact solution of the Boltzmann equation \cite{Florkowski:2014sfa} (black solid line), the full aHydro equations including the bulk degree of freedom (red dashed line), and the aHydro equations with the spheroidal bulk degree of freedom ($\tilde\Phi$) set to zero at all times (blue dot-dashed line).  For both panels we used $m=$ 1 GeV and $\tau_{\rm eq}$ = 0.5 fm/c.  In the top panels we fixed the initial spheroidal anisotropy parameter $\xi_0=0$ and in the bottom panels we chose $\xi_0 = 100$.  For the bulk initial condition we take $\tilde\Phi_0 = 0$ since this is consistent with the spheroidal initial condition assumed in the exact solution.

\begin{figure}[t]
\centerline{\includegraphics[angle=0,width=0.7\textwidth]{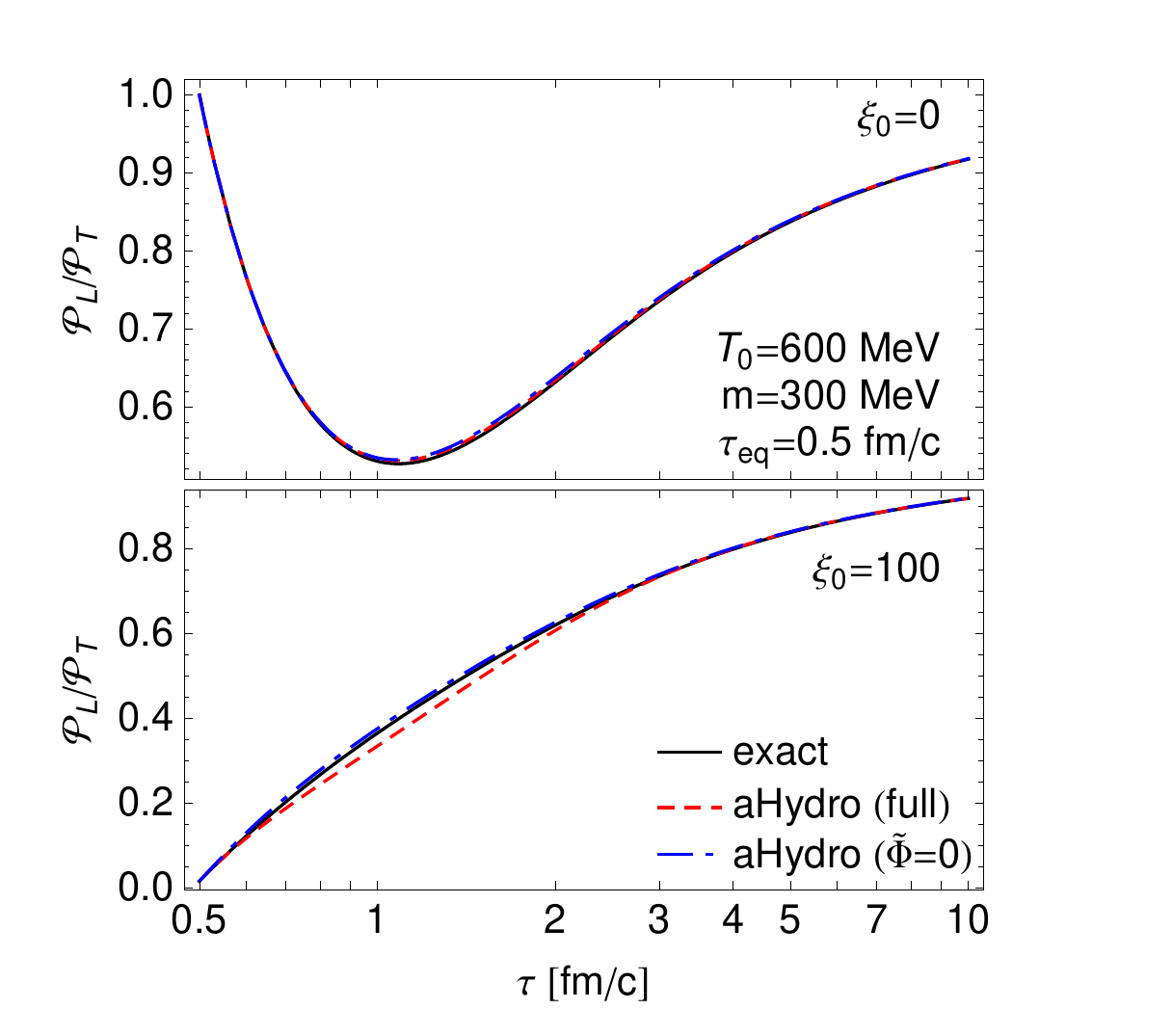}}
\caption{(Color online) Proper-time evolution of ${\cal P}_L/{\cal P}_T$. Parameters and descriptions are the same as in Fig.~\ref{fig:PL2PT_1000} except here we take $m=300$ MeV.}
\label{fig:PL2PT_300}
\end{figure}

\begin{figure}[t]
\centerline{\includegraphics[angle=0,width=0.7\textwidth]{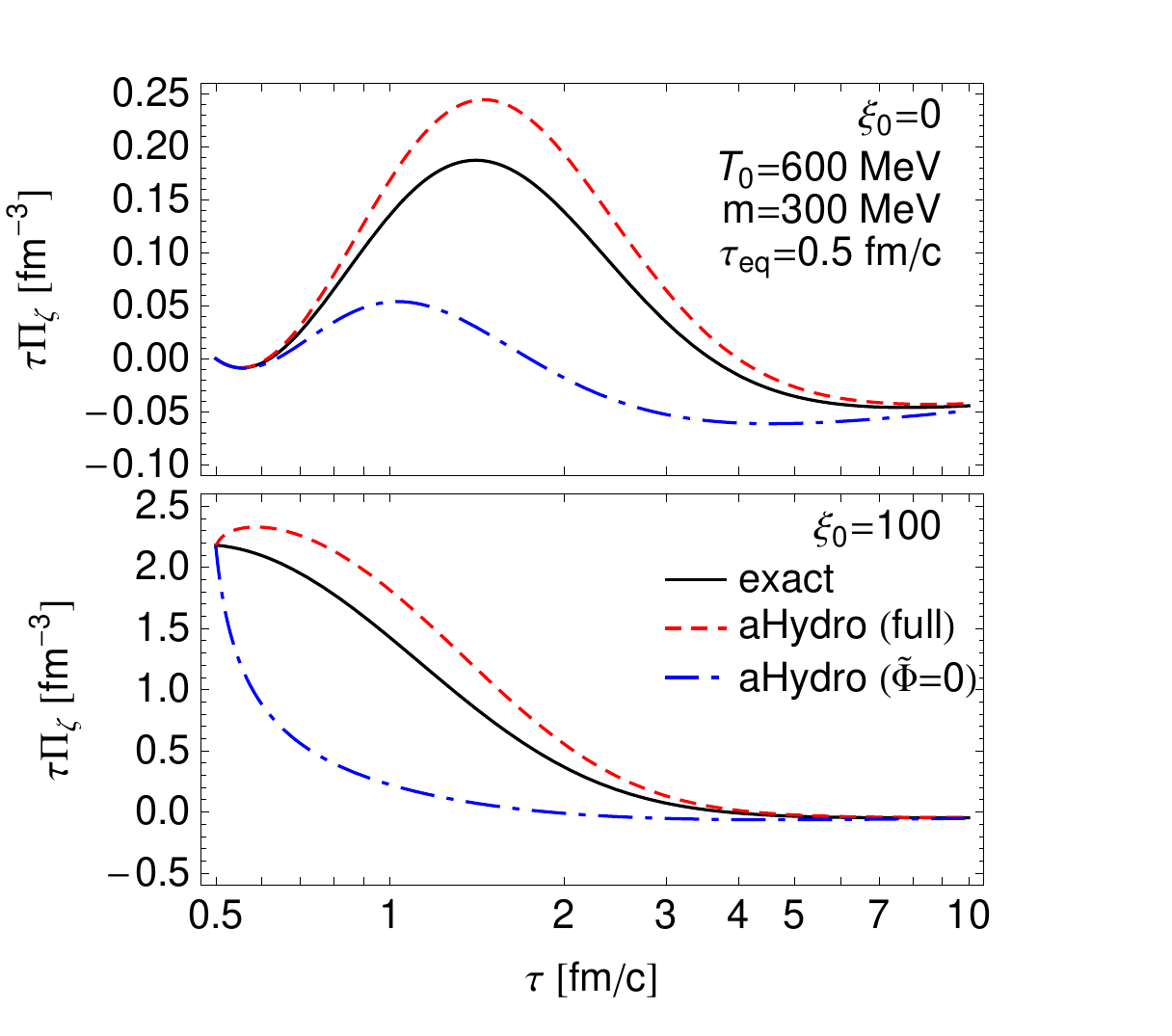}}
\caption{(Color online) Proper-time evolution of the bulk pressure.  Parameters and descriptions are the same as in Fig.~\ref{fig:PL2PT_300}.}
\label{fig:bulk_300}
\end{figure}

Considering first Fig.~\ref{fig:bulk_1000}, we see that allowing for the bulk degree of freedom significantly improves agreement between aHydro and the exact solution.  The equations derived by us previously \cite{Florkowski:2014bba} correspond to the assumption that $\tilde\Phi=0$ at all proper times.  As we can see from this figure, if this assumption is made (blue dot-dashed line) the agreement with the exact solution is quite poor.  Alternatively, one could assume that the ellipsoidal bulk parameter $\Phi=0$ at all times.  We do not show this case, because it is vastly inferior and does not even reproduce the late-time dynamics of the system.  Turning our attention to Fig.~\ref{fig:PL2PT_1000}, we see from the top panel that for a system which is initially isotropic, the different prescriptions seem to give nearly identical results for ${\cal P}_L/{\cal P}_T$.  However, if the initial pressure anisotropy is large (bottom panel), then solution of the full aHydro equations including the bulk degree of freedom seems to be further away from the exact solution.  It seems that, within the framework advocated here, it not possible to improve the agreement with the exact solutions for the bulk pressure without causing some discrepancy in the pressure anisotropy.  Since the number of parameters we used to describe the system is quite small, this may not be surprising, but it is still worrisome that we do not see uniform convergence towards the exact result in all bulk observables. 

We now consider a somewhat lower mass as an additional check of the performance of the aHydro equations obtained herein.  In Figs.~\ref{fig:PL2PT_300} and \ref{fig:bulk_300} we plot the proper-time evolution of ${\cal P}_L/{\cal P}_T$ and the bulk pressure $\Pi_\zeta$, respectively.  The parameters and descriptions are the same as Figs.~\ref{fig:PL2PT_1000} and \ref{fig:bulk_1000}, except for these figures we take $m =$ 300 MeV.  These figures once again show that including the bulk degree of freedom improves agreement between aHydro and the exact solution for the bulk pressure; however, including the bulk degree of freedom seems to cause a somewhat poorer agreement with the pressure anisotropy when the system has a large initial momentum-space anisotropy.

\begin{figure}[t]
\centerline{\includegraphics[angle=0,width=0.7\textwidth]{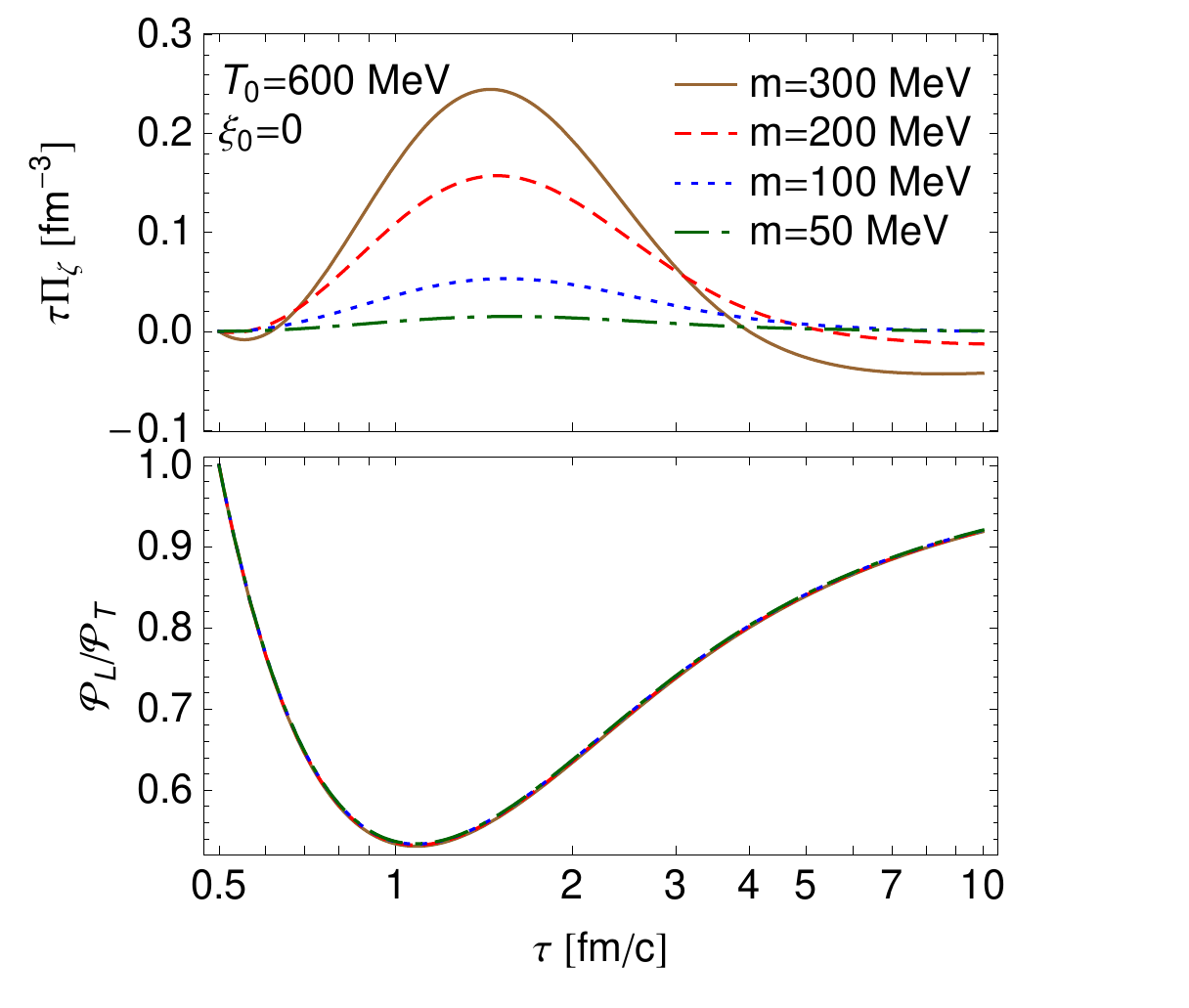}}
\caption{(Color online) Bulk pressure and pressure anisotropy as a function of proper time.  All curves correspond to the aHydro evolution including the bulk degree of freedom.  We took $\xi_0=0$ with all other the initial conditions and parameters being the same as in previous figures.}
\label{fig:bulk_varm}
\end{figure}

As our final numerical result, in Fig.~\ref{fig:bulk_varm} we plot the bulk pressure (\ref{eq:PIkz}) as a function of proper time for different assumed particle masses ranging from $m=300$ MeV down to $m=50$ MeV.  Except for the masses, the parameters, descriptions, and initial conditions are the same as in the preceding figures.  As we can see from the top panel of Fig.~\ref{fig:bulk_varm}, as one lowers the mass, the bulk pressure goes to zero as it should.  From the bottom panel we learn that there is very little dependence of the pressure anisotropy on the assumed mass of the particles.

\section{Conclusions and Outlook}
\label{sect:conc}

In this paper we have extended the treatment of Tinti and Florkowski \cite{Tinti:2013vba} to include an explicit bulk degree of freedom.  This was done by introducing a general form for the anisotropy tensor $\Xi^{\mu\nu}$ and then decomposing it into components parallel to and orthogonal to the fluid four-velocity.  We then further decomposed the orthogonal piece into traceless and traceful components analogously to how the viscous tensor is decomposed in standard relativistic viscous hydrodynamics.  Using this as a starting point, we then derived explicit expressions for the number density, energy density, and pressures for a massive anisotropic gas.  We then proceeded to take moments of the Boltzmann equation in the relaxation time approximation.  Restricting ourselves to boost-invariant and cylindrically symmetric systems we obtained the full set of dynamical equations necessary to evolve the effective temperature, momentum-space anisotropies, and the bulk degree of freedom.  

In order to test the efficacy of the approach, we then considered a transversally homogeneous system which reduces the system to 0+1d.  For such a boost-invariant and transversally homogeneous system of massive particles it is possible to solve the relaxation time approximation Boltzmann equation exactly \cite{Florkowski:2014sfa}.  Our comparisons of aHydro with the exact solution showed that adding the bulk degree of freedom improves agreement between aHydro and the exact results for the bulk pressure.  However, including this degree of freedom seems to cause some small early-time discrepancy with the pressure anisotropy evolution when the system is assumed to have a large initial momentum-space anisotropy.

On the formal side, an important result of this work concerns the question of how to select which moments of the Boltzmann equation to use for the evolution of the microscopic parameters.  Even in the case of a 0+1d system, with the addition of the bulk degree of freedom, it is not obvious a priori which moment, either zeroth moment or the $uu$-projection of the second moment, should be used as the additional equation of motion.  We demonstrated herein that in general, for a system of massive particles, the zeroth moment and $uu$-projections of the second moment of the Boltzmann equation give the same dynamical equation.  As a consequence, there is less ambiguity about how to proceed in the case of a 0+1d system.  If one considers a 1+1d boost-invariant cylindrically symmetric system there are two more equation than the number of unknowns.  The prescription of Tinti and Florkowski was to disregard the equation generated by the $ux$-projection of the second moment and the sum of the $\Theta_i$ equations, which seems to work in practice; however, it would be nice to have a more firm physics justification for this procedure.  To us, the mismatch in number of equations and parameters suggests that for a 1+1d system one can introduce an additional parameter in the ansatz for the one-particle distribution; however, it is unclear at this moment in time what additional physics parameters are required/well-motivated.

Looking forward, despite the progress reported here, there are still important open questions to be addressed.  The first and foremost question in our minds concerns the massless limit of the equations obtained herein.  In this limit the bulk degree of freedom should be irrelevant, which is evidenced by our numerical results (in that the bulk pressure goes to zero negating the need for this degree of freedom).  However, in terms of the microscopic parameters $\xi_i$ and $\Phi$ it is not obvious to us at this moment that the system of equations obtained herein reduces smoothly to the system of equations obtained by Tinti and Florkowski.  It is straightforward to show that the first and second moment equations become the same as those obtained by Tinti and Florkowski, however, the zeroth moment equation remains part of the system of equations even in the massless limit and one then has an overdetermined system.  In their approach Tinti and Florkowski disregarded the zeroth moment equation, so it is not clear to us how our equations and theirs can be smoothly connected.  

Another important open question raised by this work concerns how to simultaneously improve the description of the pressure anisotropy and bulk pressure.  In order to get better agreement with the exact solutions for the pressure anisotropy and bulk pressure, particularly at early times, it seems that one needs to account for non-ellipsoidal components of the one-particle distribution function.  This can be done using methods similar to Ref.~\cite{Bazow:2013ifa}; however, the anisotropic background would now be much more complicated.  It may be more efficient in the end to linearize around the spheroidal background and include the ellipsoidal and bulk corrections perturbatively since in this case many of the integrals (analogs of the ${\cal H}_3$ functions used herein) can be evaluated analytically.  

Finally, we note that in this work we have considered a system of particles with fixed masses which means that the equation of state is fixed.  Looking forward, within the kinetic approach one would like to have a way to implement a realistic equation of state that can reasonably reproduce the lattice equation of state.  One possibility is to use a quasiparticle approach as proposed by Romatschke~\cite{Romatschke:2011qp}.  It has been shown recently that quasiparticle-inspired HTLpt resummations of the QCD equation of state do a very good job in reproducing lattice data for the equation of state down to temperatures on the order of 200 - 300 MeV \cite{Andersen:2010wu,Andersen:2011sf,Haque:2013sja,Haque:2014rua}.  At low temperatures, massive hadron resonance gas approaches seem to work well \cite{Chojnacki:2007jc,Huovinen:2009yb}.  In the approach advocated by Romatschke, one uses a phenomenological quasiparticle model with a simple temperature-dependent (scalar) mass which is fitted to require that the lattice equation of state is reproduced.  One complication, however, is that by making the mass temperature-dependent, one violates thermodynamic consistency which requires the introduction of an additional scalar contribution to the energy-momentum tensor which can be used to enforce thermodynamic consistency. This will modify the dynamical equations presented here since there will be terms coming from the spacetime variation of this additional scalar term entering the first moment of the Boltzmann equation in addition to terms coming from the spacetime variation of the mass itself.  We leave this, and the other open questions mentioned above, for future work.

\acknowledgments{
R.R. was supported by Polish National Science Center grant No.~DEC-2012/07/D/ST2/02125, the 
Foundation for Polish Science, and U.S.~DOE Grant No.~DE-SC0004104.   M.S. was 
supported in part by U.S.~DOE Grant No.~DE-SC0004104.
}

\appendix

\section{The $\cal H$ functions}
\label{app:hfuncs}

In this appendix we collect expressions and limiting cases for the various ${\cal H}$ functions that appear in the expressions for the energy density, pressures, and dynamical equations.  In each subsection we present the general form, the spheroidal form, and near-isotropy ($\alpha_i \sim 1$) expressions in both the small- and large-$\hat{m}$ limits.

\subsection{Energy Density Integrals: ${\cal H}_3$ and $\tilde{\cal H}_3$}
\label{app:h3}

When computing the energy density using Eq.~(\ref{eq:edensint}) one obtains the following integral that, for general values of the parameters, must be computed numerically
\be
{\cal H}_3({\boldsymbol\xi},\Phi,\hat{m}) \equiv  \tilde{N} \alpha_x \alpha_y
\int_0^{2\pi} d\phi \, \alpha_\perp^2 \int_0^\infty d\hat{p} \, \hat{p}^3  f_{\rm iso}\!\left(\!\sqrt{\hat{p}^2 + \hat{m}^2}\right) {\cal H}_2\!\left(\frac{\alpha_z}{\alpha_\perp},\frac{\hat{m}}{\alpha_\perp \hat{p}} \right) ,
\label{eq:h3gen}
\ee
\checked{rs}
with $\alpha_\perp^2 \equiv \alpha_x^2 \cos^2\phi + \alpha_y^2 \sin^2\phi$, $\hat{m} \equiv m/\lambda$, $\hat{p} = |\hat{\bf p}|$, and 
\ba
 {\cal H}_2(y,z) &\equiv&
y \int_{-1}^1 d(\cos\theta)  \; 
\sqrt{y^2 \cos^2\theta + \sin^2\theta + z^2}
\nonumber \\
&=& \frac{y}{\sqrt{y^2-1}} \left( (1+z^2)
\tanh^{-1} \sqrt{\frac{y^2-1}{y^2+z^2}} + \sqrt{(y^2+z^2)(y^2-1)} \, \right).
\label{eq:H2}
\ea
\checked{rs}
For a transversally homogeneous system one has $\alpha_x = \alpha_y$ such that $\alpha_\perp = \alpha_x$ and we obtain
\be
\tilde{\cal H}_3({\boldsymbol \xi},\Phi,\hat{m}) \equiv  2 \pi \tilde{N} \alpha_x^4
\int_0^\infty d\hat{p} \, \hat{p}^3  f_{\rm iso}\!\left(\!\sqrt{\hat{p}^2 + \hat{m}^2}\right) {\cal H}_2\!\left(\frac{\alpha_z}{\alpha_x},\frac{\hat{m}}{\alpha_x\hat{p}} \right) .
\label{eq:h3tilde}
\ee
\checked{rs}

If the system is approximately isotropic ($\alpha_i \simeq 1$) one can compute the integrals above
analytically in a systematic expansion in $\delta_i \equiv \alpha_i - 1$, however, beyond leading
order in this expansion yields rather complex expressions involving generalized hypergeometric functions.  These expressions simplify considerably in the limits of either small or large $\hat{m}$.  The limit of small $\hat{m}$ is relevant to understanding the early-time dynamics and the limit of large $\hat{m}$ is relevant to the understanding of the late-time dynamics.  Taking $\delta_i \sim \epsilon$ and expanding to order $\epsilon$ in the small-$\hat{m}$ limit one obtains 
\be
\lim_{\hat{m} \rightarrow 0}
\lim_{\alpha_i \rightarrow 1}
{\cal H}_3 \simeq 2 \pi \tilde{N} \left[ 4 \left( 4 \sum_i \alpha_i - 9 \right) - \left( 2 \sum_i \alpha_i - 5 \right) \hat{m}^2 \right]  ,
\label{eq:h3eqsmallm}
\ee
\checked{rs}
and in the large-$\hat{m}$ limit one obtains
\be
\lim_{\hat{m} \rightarrow \infty}
\lim_{\alpha_i \rightarrow 1}
{\cal H}_3 \simeq (2 \pi)^{3/2} \tilde{N} \hat{m}^{5/2} e^{-\hat{m}} 
\left( \sum_i \alpha_i - 2 \right) ,
\label{eq:h3eqlargem}
\ee
\checked{rs}
where in both cases we have assumed $f_{\rm iso}$ is a Boltzmann distribution.

\subsection{Transverse Pressure Integrals: ${\cal H}_{3T}$ and $\tilde{\cal H}_{3T}$}
\label{app:h3t}

When computing the transverse pressure using Eq.~(\ref{eq:ptint}) one obtains the following integral that, for general values of the parameters, must be computed numerically
\be
{\cal H}_{3T}({\boldsymbol\xi},\Phi,\hat{m}) \equiv  \frac{1}{2} \tilde{N}\alpha_x \alpha_y
\int_0^{2\pi} d\phi \,\alpha_\perp^2 \int_0^\infty d\hat{p} \, \hat{p}^3  f_{\rm iso}\!\left(\!\sqrt{\hat{p}^2 + \hat{m}^2}\right) {\cal H}_{2T}\!\left(\frac{\alpha_z}{\alpha_\perp},\frac{\hat{m}}{\alpha_\perp \hat{p}} \right) ,
\label{eq:h3tgen}
\ee
\checked{rs}
and
\ba
{\cal H}_{2T}(y,z) &\equiv&
 y \, 
\int\limits_{-1}^1 \frac{d(\cos\theta) \sin^2\theta }{
\, \sqrt{y^2 \cos^2\theta + \sin^2\theta + z^2}} 
\label{eq:H2T} \nonumber \\
&=& \frac{y}{(y^2-1)^{3/2}}
\left[\left(z^2+2y^2-1\right) 
\tanh^{-1}\sqrt{\frac{y^2-1}{y^2+z^2}}
-\sqrt{(y^2-1)(y^2+z^2)} \right]. \hspace{1cm}
\ea
\checked{rs}
For a transversally homogeneous system one has $\alpha_x = \alpha_y$ such that $\alpha_\perp = \alpha_x$ and we obtain
\be
\tilde{\cal H}_{3T}({\boldsymbol \xi},\Phi,\hat{m}) \equiv  \pi \tilde{N} \alpha_x^4
\int_0^\infty d\hat{p} \, \hat{p}^3  f_{\rm iso}\!\left(\!\sqrt{\hat{p}^2 + \hat{m}^2}\right) {\cal H}_{2T}\!\left(\frac{\alpha_z}{\alpha_x},\frac{\hat{m}}{\alpha_x\hat{p}} \right) .
\label{eq:h3ttilde}
\ee
\checked{rs}

Taking $\delta_i \equiv \alpha_i - 1 \sim \epsilon$ and expanding to order $\epsilon$ in the small-$\hat{m}$ limit one obtains 
\be
\lim_{\hat{m} \rightarrow 0}
\lim_{\alpha_i \rightarrow 1}
{\cal H}_{3T} \simeq 2 \pi \tilde{N} \left[ \frac{4}{5} \left( 8 \alpha_x + 8 \alpha_y + 4 \alpha_z - 15 \right) - \frac{1}{3} \left( 4 \alpha_x + 4 \alpha_y + 2 \alpha_z - 7 \right) \hat{m}^2 \right] ,
\label{eq:h3teqsmallm}
\ee
\checked{rs}
and in the large-$\hat{m}$ limit one obtains
\be
\lim_{\hat{m} \rightarrow \infty}
\lim_{\alpha_i \rightarrow 1}
{\cal H}_{3T} \simeq (2 \pi)^{3/2} \tilde{N} \hat{m}^{3/2} e^{-\hat{m}} 
\left( 2 \alpha_x + 2 \alpha_y + \alpha_z - 4 \right) ,
\label{eq:h3teqlargem}
\ee
\checked{rs}
where once again we have assumed $f_{\rm iso}$ is a Boltzmann distribution.

\subsection{Longitudinal Pressure Integrals: ${\cal H}_{3L}$ and $\tilde{\cal H}_{3L}$}
\label{app:h3l}

When computing the longitudinal pressure using Eq.~(\ref{eq:plint}) one obtains the following integral that, for general values of the parameters, must be computed numerically
\be
{\cal H}_{3L}({\boldsymbol\xi},\Phi,\hat{m}) \equiv  \tilde{N} \alpha_x \alpha_y
\int_0^{2\pi} d\phi \, \alpha_\perp^2 \int_0^\infty d\hat{p} \, \hat{p}^3  f_{\rm iso}\!\left(\!\sqrt{\hat{p}^2 + \hat{m}^2}\right) {\cal H}_{2L}\!\left(\frac{\alpha_z}{\alpha_\perp},\frac{\hat{m}}{\alpha_\perp \hat{p}} \right) ,
\label{eq:h3lgen}
\ee
\checked{rs}
and
\ba
{\cal H}_{2L}(y,z) &=& y^3 \, 
 \int\limits_{-1}^1 \frac{ d(\cos\theta) \cos^2\theta }{\, \sqrt{y^2 \cos^2\theta + \sin^2\theta + z^2}}
 \nonumber \\
&=& \frac{y^3}{(y^2-1)^{3/2}}
\left[
\sqrt{(y^2-1)(y^2+z^2)}-(z^2+1)
\tanh^{-1}\sqrt{\frac{y^2-1}{y^2+z^2}} \,\,\right]. 
\label{eq:H2L}
\ea
\checked{rs}
For a transversally homogeneous system one has $\alpha_x = \alpha_y$ such that $\alpha_\perp = \alpha_x$ and we obtain
\be
\tilde{\cal H}_{3L}({\boldsymbol \xi},\Phi,\hat{m}) \equiv  2 \pi \tilde{N} \alpha_x^4
\int_0^\infty d\hat{p} \, \hat{p}^3  f_{\rm iso}\!\left(\!\sqrt{\hat{p}^2 + \hat{m}^2}\right) {\cal H}_{2L}\!\left(\frac{\alpha_z}{\alpha_x},\frac{\hat{m}}{\alpha_x\hat{p}} \right) .
\label{eq:h3ltilde}
\ee
\checked{rs}

Taking $\delta_i \equiv \alpha_i - 1 \sim \epsilon$ and expanding to order $\epsilon$ in the small-$\hat{m}$ limit one obtains 
\be
\lim_{\hat{m} \rightarrow 0}
\lim_{\alpha_i \rightarrow 1}
{\cal H}_{3L} \simeq 2 \pi \tilde{N} \left[ \frac{4}{5} \left( 4 \alpha_x + 4 \alpha_y + 12 \alpha_z - 15 \right) - \frac{1}{3} \left( 2 \alpha_x + 2 \alpha_y + 6 \alpha_z - 7 \right) \hat{m}^2 \right] ,
\label{eq:h3leqsmallm}
\ee
\checked{rs}
and in the large-$\hat{m}$ limit one obtains
\be
\lim_{\hat{m} \rightarrow \infty}
\lim_{\alpha_i \rightarrow 1}
{\cal H}_{3L} \simeq (2 \pi)^{3/2} \tilde{N} \hat{m}^{3/2} e^{-\hat{m}} 
\left( \alpha_x + \alpha_y + 3 \alpha_z - 4 \right) ,
\label{eq:h3leqlargem}
\ee
\checked{rs}
where once again we have assumed $f_{\rm iso}$ is a Boltzmann distribution.

\subsection{Mass derivative integrals: ${\cal H}_{3m}$ and $\tilde{\cal H}_{3m}$}
\label{app:h3m}

When evaluating Eq.~(\ref{eq:mintdef}) the following integral arises
\be
\tilde{\cal H}_{3m}({\boldsymbol\xi},\Phi,\hat{m}) \equiv  -2 \pi \tilde{N} \alpha_x^4 \hat{m}^2
\int_0^\infty d\hat{p} \, \frac{\hat{p}^3}{\sqrt{\hat{p}^2+\hat{m}^2}}  f_{\rm iso}'\left(\!\sqrt{\hat{p}^2 + \hat{m}^2}\right) {\cal H}_2\!\left(\frac{\alpha_z}{\alpha_x},\frac{\hat{m}}{\alpha_x \hat{p}} \right) ,
\label{eq:h3mtilde}
\ee
\checked{rs}
where $f_{\rm iso}'$ is the derivative of the isotropic distribution function with respect to its argument and ${\cal H}_2$ is defined in Eq.~(\ref{eq:H2}).

Taking $\delta_i \equiv \alpha_i - 1 \sim \epsilon$ and expanding to order $\epsilon$ in the small-$\hat{m}$ limit one obtains 
\be
\lim_{\hat{m} \rightarrow 0}
\lim_{\alpha_i \rightarrow 1}
{\cal H}_{3m} \simeq \frac{8}{3} \pi \tilde{N} \left( 4 \sum_i \alpha_i - 9 \right) \hat{m}^2 \, ,
\label{eq:h3meqsmallm}
\ee
\checked{rs}
and in the large-$\hat{m}$ limit one obtains
\be
\lim_{\hat{m} \rightarrow \infty}
\lim_{\alpha_i \rightarrow 1}
{\cal H}_{3m} \simeq (2 \pi)^{3/2} \tilde{N} \hat{m}^{7/2} e^{-\hat{m}} 
\left( \sum_i \alpha_i - 2 \right) ,
\label{eq:h3meqlargem}
\ee
\checked{rs}
where once again we have assumed $f_{\rm iso}$ is a Boltzmann distribution.

\subsection{Asymptotic expansions of the $\tilde\Omega$ functions}
\label{app:omega}

In the small-$\hat{m}$ limit one has
\ba
\tilde{\Omega}_L &\simeq& 4 \pi \tilde{N} \left[ \frac{8}{5} \left( 6 \alpha_x + 6 \alpha_y + 8 \alpha_z - 15 \right) - \frac{1}{3} \left( 4 \alpha_x + 4 \alpha_y + 6 \alpha_z - 11 \right) \hat{m}^2 \right] \, , 
\nonumber \\
\tilde{\Omega}_T &\simeq& 4 \pi \tilde{N} \left[ \frac{8}{5} \left( 7 \alpha_x + 7 \alpha_y + 6 \alpha_z - 15 \right) - \frac{1}{3} \left( 5 \alpha_x + 5 \alpha_y + 4 \alpha_z - 11 \right) \hat{m}^2 \right]  \, , 
\nonumber \\
\tilde{\Omega}_m &\simeq& - 4 \pi \tilde{N} \left( 2 \sum_i \alpha_i - 5 \right) \hat{m}^2 \, .
\ea
\checked{rs}
where we have assumed $f_{\rm iso}$ is a Boltzmann distribution.

In the large-$\hat{m}$ limit one has
\ba
\tilde{\Omega}_L &\simeq& \tilde{\cal H}_3 \, , 
\nonumber \\
\tilde{\Omega}_T &\simeq& \tilde{\cal H}_3 \, , 
\nonumber \\
\tilde{\Omega}_m &\simeq& - \tilde{\cal H}_{3m} \, .
\ea
\checked{rs}

\section{Explicit formulas for derivatives}
\label{app:derivs}

Here we repeat some formulas from the appendix of Ref.~\cite{Tinti:2013vba} using our notation as a reference point.
The convective derivatives $D = u^\mu \partial_\mu$ and $D_i = X^\mu_i \partial_\mu$ are
\ba
D &=& \cosh\theta_\perp \, \partial_\tau + \sinh\theta_\perp \, \partial_r \, ,
\nonumber \\
D_x &=& \sinh\theta_\perp \, \partial_\tau + \cosh\theta_\perp \, \partial_r \, ,
\nonumber \\
D_y &=& \frac{1}{r} \partial_\phi \, ,
\nonumber \\
D_z &=& \frac{1}{\tau} \partial_{\eta_\parallel} \, .
\label{eq:derivs1}
\ea
\checked{rs}
The divergences are
\ba
\partial_\mu u^\mu &=&  \cosh\theta_\perp \left( \frac{1}{\tau} + \partial_r \theta_\perp \right) 
+ \sinh\theta_\perp \left( \frac{1}{r} + \partial_\tau \theta_\perp \right) ,
\nonumber \\
\partial_\mu x^\mu &=&  \sinh\theta_\perp \left( \frac{1}{\tau} + \partial_r \theta_\perp \right) 
+ \cosh\theta_\perp \left( \frac{1}{r} + \partial_\tau \theta_\perp \right) ,
\nonumber \\
\partial_\mu y^\mu &=& 0 \, ,
\nonumber \\
\partial_\mu z^\mu &=& 0 \, .
\ea
\checked{rs}
One finds for the convective derivatives
\ba
D u^\mu &=& x^\mu D \theta_\perp \, ,
\nonumber \\
D x^\mu &=& u^\mu D \theta_\perp \, ,
\nonumber \\
D y^\mu &=& 0 \, ,
\nonumber \\
D z^\mu &=& 0 \, .
\ea
\checked{rs}
Likewise, for the directional derivatives $D_i \equiv X^\mu_i \partial_\mu$ one finds
\ba
D_x u^\mu &=& x^\mu D_x \theta_\perp \, ,
\nonumber \\
D_x x^\mu &=& u^\mu D_x \theta_\perp \, ,
\nonumber \\
D_x y^\mu &=& 0 \, ,
\nonumber \\
D_x z^\mu &=& 0 \, ,
\ea
\checked{rs}
and
\ba
D_y u^\mu &=& y^\mu \frac{\sinh\theta_\perp}{r} \, ,
\nonumber \\
D_y x^\mu &=& y^\mu \frac{\cosh\theta_\perp}{r} \, ,
\nonumber \\
D_y y^\mu &=& \frac{1}{r} \left( u^\mu \sinh\theta_\perp - x^\mu \cosh\theta_\perp \right) \, ,
\nonumber \\
D_y z^\mu &=& 0 \, ,
\ea
\checked{rs}
and
\ba
D_z u^\mu &=& z^\mu \frac{\cosh\theta_\perp}{\tau} \, ,
\nonumber \\
D_z x^\mu &=& z^\mu \frac{\sinh\theta_\perp}{\tau} \, ,
\nonumber \\
D_z y^\mu &=& 0 \, ,
\nonumber \\
D_z z^\mu &=& \frac{1}{\tau} \left( u^\mu \cosh\theta_\perp - x^\mu \sinh\theta_\perp \right) \, .
\ea
\checked{rs}

\bibliography{bulk}

\begin{thebibliography}{60}
\expandafter\ifx\csname natexlab\endcsname\relax\def\natexlab#1{#1}\fi
\expandafter\ifx\csname bibnamefont\endcsname\relax
  \def\bibnamefont#1{#1}\fi
\expandafter\ifx\csname bibfnamefont\endcsname\relax
  \def\bibfnamefont#1{#1}\fi
\expandafter\ifx\csname citenamefont\endcsname\relax
  \def\citenamefont#1{#1}\fi
\expandafter\ifx\csname url\endcsname\relax
  \def\url#1{\texttt{#1}}\fi
\expandafter\ifx\csname urlprefix\endcsname\relax\def\urlprefix{URL }\fi
\providecommand{\bibinfo}[2]{#2}
\providecommand{\eprint}[2][]{\url{#2}}

\bibitem[{\citenamefont{Huovinen et~al.}(2001)\citenamefont{Huovinen, Kolb,
  Heinz, Ruuskanen, and Voloshin}}]{Huovinen:2001cy}
\bibinfo{author}{\bibfnamefont{P.}~\bibnamefont{Huovinen}},
  \bibinfo{author}{\bibfnamefont{P.~F.} \bibnamefont{Kolb}},
  \bibinfo{author}{\bibfnamefont{U.~W.} \bibnamefont{Heinz}},
  \bibinfo{author}{\bibfnamefont{P.~V.} \bibnamefont{Ruuskanen}},
  \bibnamefont{and} \bibinfo{author}{\bibfnamefont{S.~A.}
  \bibnamefont{Voloshin}}, \bibinfo{journal}{Phys. Lett.}
  \textbf{\bibinfo{volume}{B503}}, \bibinfo{pages}{58} (\bibinfo{year}{2001}),
  \eprint{hep-ph/0101136}.

\bibitem[{\citenamefont{Hirano and Tsuda}(2002)}]{Hirano:2002ds}
\bibinfo{author}{\bibfnamefont{T.}~\bibnamefont{Hirano}} \bibnamefont{and}
  \bibinfo{author}{\bibfnamefont{K.}~\bibnamefont{Tsuda}},
  \bibinfo{journal}{Phys. Rev.} \textbf{\bibinfo{volume}{C66}},
  \bibinfo{pages}{054905} (\bibinfo{year}{2002}), \eprint{nucl-th/0205043}.

\bibitem[{\citenamefont{Kolb and Heinz}(2003)}]{Kolb:2003dz}
\bibinfo{author}{\bibfnamefont{P.~F.} \bibnamefont{Kolb}} \bibnamefont{and}
  \bibinfo{author}{\bibfnamefont{U.~W.} \bibnamefont{Heinz}},
  \bibinfo{journal}{In Hwa, R.C. (ed.) et al.: Quark gluon plasma} pp.
  \bibinfo{pages}{634--714} (\bibinfo{year}{2003}), \eprint{nucl-th/0305084}.

\bibitem[{\citenamefont{Muronga}(2002)}]{Muronga:2001zk}
\bibinfo{author}{\bibfnamefont{A.}~\bibnamefont{Muronga}},
  \bibinfo{journal}{Phys. Rev. Lett.} \textbf{\bibinfo{volume}{88}},
  \bibinfo{pages}{062302} (\bibinfo{year}{2002}), \eprint{nucl-th/0104064}.

\bibitem[{\citenamefont{Muronga}(2004)}]{Muronga:2003ta}
\bibinfo{author}{\bibfnamefont{A.}~\bibnamefont{Muronga}},
  \bibinfo{journal}{Phys. Rev.} \textbf{\bibinfo{volume}{C69}},
  \bibinfo{pages}{034903} (\bibinfo{year}{2004}), \eprint{nucl-th/0309055}.

\bibitem[{\citenamefont{Muronga and Rischke}(2004)}]{Muronga:2004sf}
\bibinfo{author}{\bibfnamefont{A.}~\bibnamefont{Muronga}} \bibnamefont{and}
  \bibinfo{author}{\bibfnamefont{D.~H.} \bibnamefont{Rischke}}
  (\bibinfo{year}{2004}), \eprint{nucl-th/0407114}.

\bibitem[{\citenamefont{Heinz et~al.}(2006)\citenamefont{Heinz, Song, and
  Chaudhuri}}]{Heinz:2005bw}
\bibinfo{author}{\bibfnamefont{U.~W.} \bibnamefont{Heinz}},
  \bibinfo{author}{\bibfnamefont{H.}~\bibnamefont{Song}}, \bibnamefont{and}
  \bibinfo{author}{\bibfnamefont{A.~K.} \bibnamefont{Chaudhuri}},
  \bibinfo{journal}{Phys.Rev.} \textbf{\bibinfo{volume}{C73}},
  \bibinfo{pages}{034904} (\bibinfo{year}{2006}), \eprint{nucl-th/0510014}.

\bibitem[{\citenamefont{Baier et~al.}(2006)\citenamefont{Baier, Romatschke, and
  Wiedemann}}]{Baier:2006um}
\bibinfo{author}{\bibfnamefont{R.}~\bibnamefont{Baier}},
  \bibinfo{author}{\bibfnamefont{P.}~\bibnamefont{Romatschke}},
  \bibnamefont{and} \bibinfo{author}{\bibfnamefont{U.~A.}
  \bibnamefont{Wiedemann}}, \bibinfo{journal}{Phys.Rev.}
  \textbf{\bibinfo{volume}{C73}}, \bibinfo{pages}{064903}
  (\bibinfo{year}{2006}), \eprint{hep-ph/0602249}.

\bibitem[{\citenamefont{Romatschke and Romatschke}(2007)}]{Romatschke:2007mq}
\bibinfo{author}{\bibfnamefont{P.}~\bibnamefont{Romatschke}} \bibnamefont{and}
  \bibinfo{author}{\bibfnamefont{U.}~\bibnamefont{Romatschke}},
  \bibinfo{journal}{Phys. Rev. Lett.} \textbf{\bibinfo{volume}{99}},
  \bibinfo{pages}{172301} (\bibinfo{year}{2007}), \eprint{0706.1522}.

\bibitem[{\citenamefont{Baier et~al.}(2008)\citenamefont{Baier, Romatschke,
  Son, Starinets, and Stephanov}}]{Baier:2007ix}
\bibinfo{author}{\bibfnamefont{R.}~\bibnamefont{Baier}},
  \bibinfo{author}{\bibfnamefont{P.}~\bibnamefont{Romatschke}},
  \bibinfo{author}{\bibfnamefont{D.~T.} \bibnamefont{Son}},
  \bibinfo{author}{\bibfnamefont{A.~O.} \bibnamefont{Starinets}},
  \bibnamefont{and} \bibinfo{author}{\bibfnamefont{M.~A.}
  \bibnamefont{Stephanov}}, \bibinfo{journal}{JHEP}
  \textbf{\bibinfo{volume}{0804}}, \bibinfo{pages}{100} (\bibinfo{year}{2008}),
  \eprint{0712.2451}.

\bibitem[{\citenamefont{Dusling and Teaney}(2008)}]{Dusling:2007gi}
\bibinfo{author}{\bibfnamefont{K.}~\bibnamefont{Dusling}} \bibnamefont{and}
  \bibinfo{author}{\bibfnamefont{D.}~\bibnamefont{Teaney}},
  \bibinfo{journal}{Phys. Rev.} \textbf{\bibinfo{volume}{C77}},
  \bibinfo{pages}{034905} (\bibinfo{year}{2008}), \eprint{0710.5932}.

\bibitem[{\citenamefont{Luzum and Romatschke}(2008)}]{Luzum:2008cw}
\bibinfo{author}{\bibfnamefont{M.}~\bibnamefont{Luzum}} \bibnamefont{and}
  \bibinfo{author}{\bibfnamefont{P.}~\bibnamefont{Romatschke}},
  \bibinfo{journal}{Phys. Rev.} \textbf{\bibinfo{volume}{C78}},
  \bibinfo{pages}{034915} (\bibinfo{year}{2008}), \eprint{0804.4015}.

\bibitem[{\citenamefont{Song and Heinz}(2009)}]{Song:2008hj}
\bibinfo{author}{\bibfnamefont{H.}~\bibnamefont{Song}} \bibnamefont{and}
  \bibinfo{author}{\bibfnamefont{U.~W.} \bibnamefont{Heinz}},
  \bibinfo{journal}{J.Phys.G} \textbf{\bibinfo{volume}{G36}},
  \bibinfo{pages}{064033} (\bibinfo{year}{2009}), \eprint{0812.4274}.

\bibitem[{\citenamefont{Heinz}(2010)}]{Heinz:2009xj}
\bibinfo{author}{\bibfnamefont{U.~W.} \bibnamefont{Heinz}},
  \bibinfo{journal}{Relativistic Heavy Ion Physics, Landolt-Boernstein New
  Series, I/23, edited by R. Stock, Springer Verlag, New York, Chap. 5}
  (\bibinfo{year}{2010}), \eprint{0901.4355}.

\bibitem[{\citenamefont{El et~al.}(2010)\citenamefont{El, Xu, and
  Greiner}}]{El:2009vj}
\bibinfo{author}{\bibfnamefont{A.}~\bibnamefont{El}},
  \bibinfo{author}{\bibfnamefont{Z.}~\bibnamefont{Xu}}, \bibnamefont{and}
  \bibinfo{author}{\bibfnamefont{C.}~\bibnamefont{Greiner}},
  \bibinfo{journal}{Phys. Rev.} \textbf{\bibinfo{volume}{C81}},
  \bibinfo{pages}{041901} (\bibinfo{year}{2010}), \eprint{0907.4500}.

\bibitem[{\citenamefont{Peralta-Ramos and
  Calzetta}(2009)}]{PeraltaRamos:2009kg}
\bibinfo{author}{\bibfnamefont{J.}~\bibnamefont{Peralta-Ramos}}
  \bibnamefont{and} \bibinfo{author}{\bibfnamefont{E.}~\bibnamefont{Calzetta}},
  \bibinfo{journal}{Phys. Rev.} \textbf{\bibinfo{volume}{D80}},
  \bibinfo{pages}{126002} (\bibinfo{year}{2009}), \eprint{0908.2646}.

\bibitem[{\citenamefont{Peralta-Ramos and
  Calzetta}(2010)}]{PeraltaRamos:2010je}
\bibinfo{author}{\bibfnamefont{J.}~\bibnamefont{Peralta-Ramos}}
  \bibnamefont{and} \bibinfo{author}{\bibfnamefont{E.}~\bibnamefont{Calzetta}},
  \bibinfo{journal}{Phys.Rev.} \textbf{\bibinfo{volume}{C82}},
  \bibinfo{pages}{054905} (\bibinfo{year}{2010}), \eprint{1003.1091}.

\bibitem[{\citenamefont{Denicol
  et~al.}(2010{\natexlab{a}})\citenamefont{Denicol, Kodama, and
  Koide}}]{Denicol:2010tr}
\bibinfo{author}{\bibfnamefont{G.}~\bibnamefont{Denicol}},
  \bibinfo{author}{\bibfnamefont{T.}~\bibnamefont{Kodama}}, \bibnamefont{and}
  \bibinfo{author}{\bibfnamefont{T.}~\bibnamefont{Koide}},
  \bibinfo{journal}{J.Phys.G} \textbf{\bibinfo{volume}{G37}},
  \bibinfo{pages}{094040} (\bibinfo{year}{2010}{\natexlab{a}}),
  \eprint{1002.2394}.

\bibitem[{\citenamefont{Denicol
  et~al.}(2010{\natexlab{b}})\citenamefont{Denicol, Koide, and
  Rischke}}]{Denicol:2010xn}
\bibinfo{author}{\bibfnamefont{G.}~\bibnamefont{Denicol}},
  \bibinfo{author}{\bibfnamefont{T.}~\bibnamefont{Koide}}, \bibnamefont{and}
  \bibinfo{author}{\bibfnamefont{D.}~\bibnamefont{Rischke}},
  \bibinfo{journal}{Phys.Rev.Lett.} \textbf{\bibinfo{volume}{105}},
  \bibinfo{pages}{162501} (\bibinfo{year}{2010}{\natexlab{b}}),
  \eprint{1004.5013}.

\bibitem[{\citenamefont{Schenke
  et~al.}(2011{\natexlab{a}})\citenamefont{Schenke, Jeon, and
  Gale}}]{Schenke:2010rr}
\bibinfo{author}{\bibfnamefont{B.}~\bibnamefont{Schenke}},
  \bibinfo{author}{\bibfnamefont{S.}~\bibnamefont{Jeon}}, \bibnamefont{and}
  \bibinfo{author}{\bibfnamefont{C.}~\bibnamefont{Gale}},
  \bibinfo{journal}{Phys.Rev.Lett.} \textbf{\bibinfo{volume}{106}},
  \bibinfo{pages}{042301} (\bibinfo{year}{2011}{\natexlab{a}}),
  \eprint{1009.3244}.

\bibitem[{\citenamefont{Schenke
  et~al.}(2011{\natexlab{b}})\citenamefont{Schenke, Jeon, and
  Gale}}]{Schenke:2011tv}
\bibinfo{author}{\bibfnamefont{B.}~\bibnamefont{Schenke}},
  \bibinfo{author}{\bibfnamefont{S.}~\bibnamefont{Jeon}}, \bibnamefont{and}
  \bibinfo{author}{\bibfnamefont{C.}~\bibnamefont{Gale}},
  \bibinfo{journal}{Phys.Lett.} \textbf{\bibinfo{volume}{B702}},
  \bibinfo{pages}{59} (\bibinfo{year}{2011}{\natexlab{b}}), \eprint{1102.0575}.

\bibitem[{\citenamefont{Bozek}(2011)}]{Bozek:2011wa}
\bibinfo{author}{\bibfnamefont{P.}~\bibnamefont{Bozek}},
  \bibinfo{journal}{Phys.Lett.} \textbf{\bibinfo{volume}{B699}},
  \bibinfo{pages}{283} (\bibinfo{year}{2011}), \eprint{1101.1791}.

\bibitem[{\citenamefont{Niemi et~al.}(2011)\citenamefont{Niemi, Denicol,
  Huovinen, Molnar, and Rischke}}]{Niemi:2011ix}
\bibinfo{author}{\bibfnamefont{H.}~\bibnamefont{Niemi}},
  \bibinfo{author}{\bibfnamefont{G.~S.} \bibnamefont{Denicol}},
  \bibinfo{author}{\bibfnamefont{P.}~\bibnamefont{Huovinen}},
  \bibinfo{author}{\bibfnamefont{E.}~\bibnamefont{Molnar}}, \bibnamefont{and}
  \bibinfo{author}{\bibfnamefont{D.~H.} \bibnamefont{Rischke}},
  \bibinfo{journal}{Phys.Rev.Lett.} \textbf{\bibinfo{volume}{106}},
  \bibinfo{pages}{212302} (\bibinfo{year}{2011}), \eprint{1101.2442}.

\bibitem[{\citenamefont{Niemi et~al.}(2012)\citenamefont{Niemi, Denicol,
  Huovinen, Moln\'ar, and Rischke}}]{Niemi:2012ry}
\bibinfo{author}{\bibfnamefont{H.}~\bibnamefont{Niemi}},
  \bibinfo{author}{\bibfnamefont{G.~S.} \bibnamefont{Denicol}},
  \bibinfo{author}{\bibfnamefont{P.}~\bibnamefont{Huovinen}},
  \bibinfo{author}{\bibfnamefont{E.}~\bibnamefont{Moln\'ar}}, \bibnamefont{and}
  \bibinfo{author}{\bibfnamefont{D.~H.} \bibnamefont{Rischke}},
  \bibinfo{journal}{Phys. Rev. C} \textbf{\bibinfo{volume}{86}},
  \bibinfo{pages}{014909} (\bibinfo{year}{2012}).

\bibitem[{\citenamefont{Bo\ifmmode~\dot{z}\else \.{z}\fi{}ek and
  Wyskiel-Piekarska}(2012)}]{Bozek:2012qs}
\bibinfo{author}{\bibfnamefont{P.}~\bibnamefont{Bo\ifmmode~\dot{z}\else
  \.{z}\fi{}ek}} \bibnamefont{and}
  \bibinfo{author}{\bibfnamefont{I.}~\bibnamefont{Wyskiel-Piekarska}},
  \bibinfo{journal}{Phys. Rev. C} \textbf{\bibinfo{volume}{85}},
  \bibinfo{pages}{064915} (\bibinfo{year}{2012}).

\bibitem[{\citenamefont{Denicol
  et~al.}(2012{\natexlab{a}})\citenamefont{Denicol, Niemi, Moln\'ar, and
  Rischke}}]{Denicol:2012cn}
\bibinfo{author}{\bibfnamefont{G.~S.} \bibnamefont{Denicol}},
  \bibinfo{author}{\bibfnamefont{H.}~\bibnamefont{Niemi}},
  \bibinfo{author}{\bibfnamefont{E.}~\bibnamefont{Moln\'ar}}, \bibnamefont{and}
  \bibinfo{author}{\bibfnamefont{D.~H.} \bibnamefont{Rischke}},
  \bibinfo{journal}{Phys. Rev. D} \textbf{\bibinfo{volume}{85}},
  \bibinfo{pages}{114047} (\bibinfo{year}{2012}{\natexlab{a}}).

\bibitem[{\citenamefont{Denicol
  et~al.}(2012{\natexlab{b}})\citenamefont{Denicol, Molnar, Niemi, and
  Rischke}}]{Denicol:2012es}
\bibinfo{author}{\bibfnamefont{G.}~\bibnamefont{Denicol}},
  \bibinfo{author}{\bibfnamefont{E.}~\bibnamefont{Molnar}},
  \bibinfo{author}{\bibfnamefont{H.}~\bibnamefont{Niemi}}, \bibnamefont{and}
  \bibinfo{author}{\bibfnamefont{D.}~\bibnamefont{Rischke}},
  \bibinfo{journal}{Eur. Phys. J. A} \textbf{\bibinfo{volume}{48}},
  \bibinfo{pages}{170} (\bibinfo{year}{2012}{\natexlab{b}}),
  \eprint{1206.1554}.

\bibitem[{\citenamefont{Peralta-Ramos and
  Calzetta}(2013)}]{PeraltaRamos:2012xk}
\bibinfo{author}{\bibfnamefont{J.}~\bibnamefont{Peralta-Ramos}}
  \bibnamefont{and} \bibinfo{author}{\bibfnamefont{E.}~\bibnamefont{Calzetta}},
  \bibinfo{journal}{Phys.Rev.} \textbf{\bibinfo{volume}{D87}},
  \bibinfo{pages}{034003} (\bibinfo{year}{2013}), \eprint{1212.0824}.

\bibitem[{\citenamefont{Calzetta}(2014)}]{Calzetta:2014hra}
\bibinfo{author}{\bibfnamefont{E.}~\bibnamefont{Calzetta}}
  (\bibinfo{year}{2014}), \eprint{1402.5278}.

\bibitem[{\citenamefont{Denicol et~al.}(2014)\citenamefont{Denicol, Jeon, and
  Gale}}]{Denicol:2014vaa}
\bibinfo{author}{\bibfnamefont{G.}~\bibnamefont{Denicol}},
  \bibinfo{author}{\bibfnamefont{S.}~\bibnamefont{Jeon}}, \bibnamefont{and}
  \bibinfo{author}{\bibfnamefont{C.}~\bibnamefont{Gale}}
  (\bibinfo{year}{2014}), \eprint{1403.0962}.

\bibitem[{\citenamefont{Kovtun et~al.}(2005)\citenamefont{Kovtun, Son, and
  Starinets}}]{Kovtun:2004de}
\bibinfo{author}{\bibfnamefont{P.}~\bibnamefont{Kovtun}},
  \bibinfo{author}{\bibfnamefont{D.}~\bibnamefont{Son}}, \bibnamefont{and}
  \bibinfo{author}{\bibfnamefont{A.}~\bibnamefont{Starinets}},
  \bibinfo{journal}{Phys.Rev.Lett.} \textbf{\bibinfo{volume}{94}},
  \bibinfo{pages}{111601} (\bibinfo{year}{2005}), \eprint{hep-th/0405231}.

\bibitem[{\citenamefont{Martinez and Strickland}(2009)}]{Martinez:2009mf}
\bibinfo{author}{\bibfnamefont{M.}~\bibnamefont{Martinez}} \bibnamefont{and}
  \bibinfo{author}{\bibfnamefont{M.}~\bibnamefont{Strickland}},
  \bibinfo{journal}{Phys. Rev.} \textbf{\bibinfo{volume}{C79}},
  \bibinfo{pages}{044903} (\bibinfo{year}{2009}), \eprint{0902.3834}.

\bibitem[{\citenamefont{Strickland}(2013)}]{Strickland:2013uga}
\bibinfo{author}{\bibfnamefont{M.}~\bibnamefont{Strickland}}
  (\bibinfo{year}{2013}), \eprint{1312.2285}.

\bibitem[{\citenamefont{Strickland}(2014)}]{Strickland:2014eua}
\bibinfo{author}{\bibfnamefont{M.}~\bibnamefont{Strickland}}
  (\bibinfo{year}{2014}), \eprint{1401.1188}.

\bibitem[{\citenamefont{Martinez and Strickland}(2010)}]{Martinez:2010sc}
\bibinfo{author}{\bibfnamefont{M.}~\bibnamefont{Martinez}} \bibnamefont{and}
  \bibinfo{author}{\bibfnamefont{M.}~\bibnamefont{Strickland}},
  \bibinfo{journal}{Nucl. Phys.} \textbf{\bibinfo{volume}{A848}},
  \bibinfo{pages}{183} (\bibinfo{year}{2010}), \eprint{1007.0889}.

\bibitem[{\citenamefont{Florkowski and Ryblewski}(2011)}]{Florkowski:2010cf}
\bibinfo{author}{\bibfnamefont{W.}~\bibnamefont{Florkowski}} \bibnamefont{and}
  \bibinfo{author}{\bibfnamefont{R.}~\bibnamefont{Ryblewski}},
  \bibinfo{journal}{Phys.Rev.} \textbf{\bibinfo{volume}{C83}},
  \bibinfo{pages}{034907} (\bibinfo{year}{2011}), \eprint{1007.0130}.

\bibitem[{\citenamefont{Ryblewski and
  Florkowski}(2011{\natexlab{a}})}]{Ryblewski:2010bs}
\bibinfo{author}{\bibfnamefont{R.}~\bibnamefont{Ryblewski}} \bibnamefont{and}
  \bibinfo{author}{\bibfnamefont{W.}~\bibnamefont{Florkowski}},
  \bibinfo{journal}{J.Phys.G} \textbf{\bibinfo{volume}{G38}},
  \bibinfo{pages}{015104} (\bibinfo{year}{2011}{\natexlab{a}}),
  \eprint{1007.4662}.

\bibitem[{\citenamefont{Martinez and Strickland}(2011)}]{Martinez:2010sd}
\bibinfo{author}{\bibfnamefont{M.}~\bibnamefont{Martinez}} \bibnamefont{and}
  \bibinfo{author}{\bibfnamefont{M.}~\bibnamefont{Strickland}},
  \bibinfo{journal}{Nucl.Phys.} \textbf{\bibinfo{volume}{A856}},
  \bibinfo{pages}{68} (\bibinfo{year}{2011}), \eprint{1011.3056}.

\bibitem[{\citenamefont{Ryblewski and
  Florkowski}(2011{\natexlab{b}})}]{Ryblewski:2011aq}
\bibinfo{author}{\bibfnamefont{R.}~\bibnamefont{Ryblewski}} \bibnamefont{and}
  \bibinfo{author}{\bibfnamefont{W.}~\bibnamefont{Florkowski}},
  \bibinfo{journal}{Eur.Phys.J.} \textbf{\bibinfo{volume}{C71}},
  \bibinfo{pages}{1761} (\bibinfo{year}{2011}{\natexlab{b}}),
  \eprint{1103.1260}.

\bibitem[{\citenamefont{Florkowski and Ryblewski}(2012)}]{Florkowski:2011jg}
\bibinfo{author}{\bibfnamefont{W.}~\bibnamefont{Florkowski}} \bibnamefont{and}
  \bibinfo{author}{\bibfnamefont{R.}~\bibnamefont{Ryblewski}},
  \bibinfo{journal}{Phys.Rev.} \textbf{\bibinfo{volume}{C85}},
  \bibinfo{pages}{044902} (\bibinfo{year}{2012}), \eprint{1111.5997}.

\bibitem[{\citenamefont{Martinez et~al.}(2012)\citenamefont{Martinez,
  Ryblewski, and Strickland}}]{Martinez:2012tu}
\bibinfo{author}{\bibfnamefont{M.}~\bibnamefont{Martinez}},
  \bibinfo{author}{\bibfnamefont{R.}~\bibnamefont{Ryblewski}},
  \bibnamefont{and}
  \bibinfo{author}{\bibfnamefont{M.}~\bibnamefont{Strickland}},
  \bibinfo{journal}{Phys.Rev.} \textbf{\bibinfo{volume}{C85}},
  \bibinfo{pages}{064913} (\bibinfo{year}{2012}), \eprint{1204.1473}.

\bibitem[{\citenamefont{Ryblewski and Florkowski}(2012)}]{Ryblewski:2012rr}
\bibinfo{author}{\bibfnamefont{R.}~\bibnamefont{Ryblewski}} \bibnamefont{and}
  \bibinfo{author}{\bibfnamefont{W.}~\bibnamefont{Florkowski}},
  \bibinfo{journal}{Phys.Rev.} \textbf{\bibinfo{volume}{C85}},
  \bibinfo{pages}{064901} (\bibinfo{year}{2012}), \eprint{1204.2624}.

\bibitem[{\citenamefont{Florkowski
  et~al.}(2013{\natexlab{a}})\citenamefont{Florkowski, Maj, Ryblewski, and
  Strickland}}]{Florkowski:2012as}
\bibinfo{author}{\bibfnamefont{W.}~\bibnamefont{Florkowski}},
  \bibinfo{author}{\bibfnamefont{R.}~\bibnamefont{Maj}},
  \bibinfo{author}{\bibfnamefont{R.}~\bibnamefont{Ryblewski}},
  \bibnamefont{and}
  \bibinfo{author}{\bibfnamefont{M.}~\bibnamefont{Strickland}},
  \bibinfo{journal}{Phys.Rev.} \textbf{\bibinfo{volume}{C87}},
  \bibinfo{pages}{034914} (\bibinfo{year}{2013}{\natexlab{a}}),
  \eprint{1209.3671}.

\bibitem[{\citenamefont{Bazow et~al.}(2013)\citenamefont{Bazow, Heinz, and
  Strickland}}]{Bazow:2013ifa}
\bibinfo{author}{\bibfnamefont{D.}~\bibnamefont{Bazow}},
  \bibinfo{author}{\bibfnamefont{U.~W.} \bibnamefont{Heinz}}, \bibnamefont{and}
  \bibinfo{author}{\bibfnamefont{M.}~\bibnamefont{Strickland}}
  (\bibinfo{year}{2013}), \eprint{1311.6720}.

\bibitem[{\citenamefont{Florkowski and Maj}(2013)}]{Florkowski:2013uqa}
\bibinfo{author}{\bibfnamefont{W.}~\bibnamefont{Florkowski}} \bibnamefont{and}
  \bibinfo{author}{\bibfnamefont{R.}~\bibnamefont{Maj}}, \bibinfo{journal}{Acta
  Phys.Polon.} \textbf{\bibinfo{volume}{B44}}, \bibinfo{pages}{2003}
  (\bibinfo{year}{2013}), \eprint{1309.2786}.

\bibitem[{\citenamefont{Tinti and Florkowski}(2014)}]{Tinti:2013vba}
\bibinfo{author}{\bibfnamefont{L.}~\bibnamefont{Tinti}} \bibnamefont{and}
  \bibinfo{author}{\bibfnamefont{W.}~\bibnamefont{Florkowski}},
  \bibinfo{journal}{Phys.Rev.} \textbf{\bibinfo{volume}{C89}},
  \bibinfo{pages}{034907} (\bibinfo{year}{2014}), \eprint{1312.6614}.

\bibitem[{\citenamefont{Florkowski and Madetko}(2014)}]{Florkowski:2014txa}
\bibinfo{author}{\bibfnamefont{W.}~\bibnamefont{Florkowski}} \bibnamefont{and}
  \bibinfo{author}{\bibfnamefont{O.}~\bibnamefont{Madetko}}
  (\bibinfo{year}{2014}), \eprint{1402.2401}.

\bibitem[{\citenamefont{Florkowski
  et~al.}(2014{\natexlab{a}})\citenamefont{Florkowski, Ryblewski, Strickland,
  and Tinti}}]{Florkowski:2014bba}
\bibinfo{author}{\bibfnamefont{W.}~\bibnamefont{Florkowski}},
  \bibinfo{author}{\bibfnamefont{R.}~\bibnamefont{Ryblewski}},
  \bibinfo{author}{\bibfnamefont{M.}~\bibnamefont{Strickland}},
  \bibnamefont{and} \bibinfo{author}{\bibfnamefont{L.}~\bibnamefont{Tinti}}
  (\bibinfo{year}{2014}{\natexlab{a}}), \eprint{1403.1223}.

\bibitem[{\citenamefont{Song}(2009)}]{Song:2009gc}
\bibinfo{author}{\bibfnamefont{H.}~\bibnamefont{Song}} (\bibinfo{year}{2009}),
  \eprint{0908.3656}.

\bibitem[{\citenamefont{Romatschke and Strickland}(2003)}]{Romatschke:2003ms}
\bibinfo{author}{\bibfnamefont{P.}~\bibnamefont{Romatschke}} \bibnamefont{and}
  \bibinfo{author}{\bibfnamefont{M.}~\bibnamefont{Strickland}},
  \bibinfo{journal}{Phys. Rev.} \textbf{\bibinfo{volume}{D68}},
  \bibinfo{pages}{036004} (\bibinfo{year}{2003}), \eprint{hep-ph/0304092}.

\bibitem[{\citenamefont{Florkowski
  et~al.}(2013{\natexlab{b}})\citenamefont{Florkowski, Ryblewski, and
  Strickland}}]{Florkowski:2013lza}
\bibinfo{author}{\bibfnamefont{W.}~\bibnamefont{Florkowski}},
  \bibinfo{author}{\bibfnamefont{R.}~\bibnamefont{Ryblewski}},
  \bibnamefont{and}
  \bibinfo{author}{\bibfnamefont{M.}~\bibnamefont{Strickland}},
  \bibinfo{journal}{Nucl.Phys.} \textbf{\bibinfo{volume}{A916}},
  \bibinfo{pages}{249} (\bibinfo{year}{2013}{\natexlab{b}}),
  \eprint{1304.0665}.

\bibitem[{\citenamefont{Florkowski
  et~al.}(2013{\natexlab{c}})\citenamefont{Florkowski, Ryblewski, and
  Strickland}}]{Florkowski:2013lya}
\bibinfo{author}{\bibfnamefont{W.}~\bibnamefont{Florkowski}},
  \bibinfo{author}{\bibfnamefont{R.}~\bibnamefont{Ryblewski}},
  \bibnamefont{and}
  \bibinfo{author}{\bibfnamefont{M.}~\bibnamefont{Strickland}},
  \bibinfo{journal}{Phys. Rev.} \textbf{\bibinfo{volume}{C88}},
  \bibinfo{pages}{024903} (\bibinfo{year}{2013}{\natexlab{c}}),
  \eprint{1305.7234}.

\bibitem[{\citenamefont{Florkowski
  et~al.}(2014{\natexlab{b}})\citenamefont{Florkowski, Maksymiuk, Ryblewski,
  and Strickland}}]{Florkowski:2014sfa}
\bibinfo{author}{\bibfnamefont{W.}~\bibnamefont{Florkowski}},
  \bibinfo{author}{\bibfnamefont{E.}~\bibnamefont{Maksymiuk}},
  \bibinfo{author}{\bibfnamefont{R.}~\bibnamefont{Ryblewski}},
  \bibnamefont{and}
  \bibinfo{author}{\bibfnamefont{M.}~\bibnamefont{Strickland}}
  (\bibinfo{year}{2014}{\natexlab{b}}), \eprint{1402.7348}.

\bibitem[{\citenamefont{Romatschke}(2012)}]{Romatschke:2011qp}
\bibinfo{author}{\bibfnamefont{P.}~\bibnamefont{Romatschke}},
  \bibinfo{journal}{Phys.Rev.} \textbf{\bibinfo{volume}{D85}},
  \bibinfo{pages}{065012} (\bibinfo{year}{2012}), \eprint{1108.5561}.

\bibitem[{\citenamefont{Andersen
  et~al.}(2011{\natexlab{a}})\citenamefont{Andersen, Leganger, Strickland, and
  Su}}]{Andersen:2010wu}
\bibinfo{author}{\bibfnamefont{J.~O.} \bibnamefont{Andersen}},
  \bibinfo{author}{\bibfnamefont{L.~E.} \bibnamefont{Leganger}},
  \bibinfo{author}{\bibfnamefont{M.}~\bibnamefont{Strickland}},
  \bibnamefont{and} \bibinfo{author}{\bibfnamefont{N.}~\bibnamefont{Su}},
  \bibinfo{journal}{Phys.Lett.} \textbf{\bibinfo{volume}{B696}},
  \bibinfo{pages}{468} (\bibinfo{year}{2011}{\natexlab{a}}),
  \eprint{1009.4644}.

\bibitem[{\citenamefont{Andersen
  et~al.}(2011{\natexlab{b}})\citenamefont{Andersen, Leganger, Strickland, and
  Su}}]{Andersen:2011sf}
\bibinfo{author}{\bibfnamefont{J.~O.} \bibnamefont{Andersen}},
  \bibinfo{author}{\bibfnamefont{L.~E.} \bibnamefont{Leganger}},
  \bibinfo{author}{\bibfnamefont{M.}~\bibnamefont{Strickland}},
  \bibnamefont{and} \bibinfo{author}{\bibfnamefont{N.}~\bibnamefont{Su}},
  \bibinfo{journal}{JHEP} \textbf{\bibinfo{volume}{1108}}, \bibinfo{pages}{053}
  (\bibinfo{year}{2011}{\natexlab{b}}), \eprint{1103.2528}.

\bibitem[{\citenamefont{Haque et~al.}(2014{\natexlab{a}})\citenamefont{Haque,
  Andersen, Mustafa, Strickland, and Su}}]{Haque:2013sja}
\bibinfo{author}{\bibfnamefont{N.}~\bibnamefont{Haque}},
  \bibinfo{author}{\bibfnamefont{J.~O.} \bibnamefont{Andersen}},
  \bibinfo{author}{\bibfnamefont{M.~G.} \bibnamefont{Mustafa}},
  \bibinfo{author}{\bibfnamefont{M.}~\bibnamefont{Strickland}},
  \bibnamefont{and} \bibinfo{author}{\bibfnamefont{N.}~\bibnamefont{Su}},
  \bibinfo{journal}{Phys.Rev.} \textbf{\bibinfo{volume}{D89}},
  \bibinfo{pages}{061701} (\bibinfo{year}{2014}{\natexlab{a}}),
  \eprint{1309.3968}.

\bibitem[{\citenamefont{Haque et~al.}(2014{\natexlab{b}})\citenamefont{Haque,
  Bandyopadhyay, Andersen, Mustafa, Strickland et~al.}}]{Haque:2014rua}
\bibinfo{author}{\bibfnamefont{N.}~\bibnamefont{Haque}},
  \bibinfo{author}{\bibfnamefont{A.}~\bibnamefont{Bandyopadhyay}},
  \bibinfo{author}{\bibfnamefont{J.~O.} \bibnamefont{Andersen}},
  \bibinfo{author}{\bibfnamefont{M.~G.} \bibnamefont{Mustafa}},
  \bibinfo{author}{\bibfnamefont{M.}~\bibnamefont{Strickland}},
  \bibnamefont{et~al.}, \bibinfo{journal}{JHEP}
  \textbf{\bibinfo{volume}{1405}}, \bibinfo{pages}{027}
  (\bibinfo{year}{2014}{\natexlab{b}}), \eprint{1402.6907}.

\bibitem[{\citenamefont{Chojnacki and Florkowski}(2007)}]{Chojnacki:2007jc}
\bibinfo{author}{\bibfnamefont{M.}~\bibnamefont{Chojnacki}} \bibnamefont{and}
  \bibinfo{author}{\bibfnamefont{W.}~\bibnamefont{Florkowski}},
  \bibinfo{journal}{Acta Phys.Polon.} \textbf{\bibinfo{volume}{B38}},
  \bibinfo{pages}{3249} (\bibinfo{year}{2007}), \eprint{nucl-th/0702030}.

\bibitem[{\citenamefont{Huovinen and Petreczky}(2010)}]{Huovinen:2009yb}
\bibinfo{author}{\bibfnamefont{P.}~\bibnamefont{Huovinen}} \bibnamefont{and}
  \bibinfo{author}{\bibfnamefont{P.}~\bibnamefont{Petreczky}},
  \bibinfo{journal}{Nucl.Phys.} \textbf{\bibinfo{volume}{A837}},
  \bibinfo{pages}{26} (\bibinfo{year}{2010}), \eprint{0912.2541}.

\end{thebibliography}

\end{document}